\documentclass[journal]{IEEEtran}
\usepackage{subfigure}
\usepackage{graphicx}
\usepackage{amssymb,amsmath,bm}
\usepackage{textcomp}

\usepackage{epsfig,epstopdf}
\usepackage{stfloats,multirow,booktabs,color,ifpdf,float,array}

\usepackage{amsmath,graphicx}

\usepackage{bm}
\usepackage{amsfonts}
\usepackage{psfrag}
\usepackage{algorithm,algorithmic}
\usepackage{tabu}

\usepackage{booktabs}

\begin{document}
%
\title{Bayesian Learning for \\ Deep Neural Network Adaptation}
%
%
%

\author{Xurong Xie,
        ~~Xunying Liu, \IEEEmembership{Member,~IEEE},
        ~~Tan Lee, \IEEEmembership{Member,~IEEE},
        ~~Lan Wang, \IEEEmembership{Member,~IEEE}
\thanks{Corresponding authors: Xunying Liu and Lan Wang.}
\thanks{Xurong Xie is with Shenzhen Institutes of Advanced Technology, Chinese Academy of Sciences, Shenzhen, China, and he was with Chinese University of Hong Kong, Hong Kong, China when he worked on this work,
e-mail: xr.xie@siat.ac.cn}
\thanks{Xunying Liu, and Tan Lee are with Chinese University of Hong Kong, Hong Kong, China,
e-mail: xyliu@se.cuhk.edu.hk, tanlee@ee.cuhk.edu.hk.}
\thanks{Lan Wang is with Shenzhen Institutes of Advanced Technology, Chinese Academy of Sciences, Shenzhen, China,
e-mail: lan.wang@siat.ac.cn.}
}

%
%

\markboth{Journal of \LaTeX\ Class Files,~Vol.~14, No.~8, August~2015}%
{Shell \MakeLowercase{\textit{et al.}}: Bare Demo of IEEEtran.cls for IEEE Journals}
%



\maketitle

\begin{abstract}
(*This manuscript is an updated version of the article published in TASLP, with appendix of released codes (in https://github.com/XIEXurong/kaldi.git) based on Kaldi toolkit, and appendix of updated experiment results on the \emph{Switchboard}, \emph{HKUST}, and \emph{DementiaBank Pitt} corpuses. Please still cite the published article in the TASLP, DOI: 10.1109/TASLP.2021.3084072.)

A key task for speech recognition systems is to reduce the mismatch between training and evaluation data that is often attributable to speaker differences. Speaker adaptation techniques play a vital role to reduce the mismatch. Model-based speaker adaptation approaches often require sufficient amounts of target speaker data to ensure robustness. When the amount of speaker level data is limited, speaker adaptation is prone to overfitting and poor generalization. To address the issue, this paper proposes a full Bayesian learning based DNN speaker adaptation framework to model speaker-dependent (SD) parameter uncertainty given limited speaker specific adaptation data. This framework is investigated in three forms of model based DNN adaptation techniques: Bayesian learning of hidden unit contributions (BLHUC), Bayesian parameterized activation functions (BPAct), and Bayesian hidden unit bias vectors (BHUB). In the three methods, deterministic SD parameters are replaced by latent variable posterior distributions for each speaker, whose parameters are efficiently estimated using a variational inference based approach. Experiments conducted on 300-hour speed perturbed \emph{Switchboard} corpus trained LF-MMI  TDNN/CNN-TDNN systems suggest the proposed Bayesian adaptation approaches consistently outperform the deterministic adaptation on the \emph{NIST Hub5'00} and \emph{RT03} evaluation sets.
When using only the first five utterances from each speaker as adaptation data, significant word error rate reductions up to 1.4\% absolute (7.2\% relative) were obtained on the \emph{CallHome} subset. The efficacy of the proposed Bayesian adaptation techniques is further demonstrated in a comparison against the state-of-the-art performance obtained on the same task using the most recent systems reported in the literature. 
\end{abstract}


%
\IEEEpeerreviewmaketitle


\section{Introduction}
\label{sec:intro}


In recent decades the performance of automatic speech recognition (ASR) systems have been significantly improved with the successful application of deep learning techniques. In the conventional deep neural network-hidden Markov models (DNN-HMMs) based hybrid ASR systems~\cite{dahl2011context,bourlard2012connectionist}, artificial neural networks are used to estimate the conditional probabilities of HMM states given acoustic features. In these systems, word sequences are transformed into phoneme sequences using a pronunciation lexicon. The resulting phoneme sequences are modeled by HMMs representing phonetic units. In order to sufficiently model long range temporal contexts and spectral characteristics of continuous speech, advanced forms of DNNs including convolutional neural networks (CNNs)~\cite{abdel2012applying,abdel2013exploring}, time-delay neural networks (TDNNs)~\cite{peddinti2015time}, recurrent neural networks (RNNs)~\cite{deng2014sequence} and their long short-term memory variants~\cite{sak-etal-is2014} as well as self-attention~\cite{wang2020transformer} are widely used in current hybrid DNN-HMM systems. In order to reduce the mismatch between the conventional frame level cross-entropy (CE) based loss function and recognition error rate normally measured at word level over a sentence, sequence level training criteria based on maximum mutual information (MMI)~\cite{kapadia1993mmi}, minimum phone error (MPE)~\cite{povey2002minimum}, segmental minimum Bayes-risk (sMBR)~\cite{vesely2013sequence}, or lattice-free MMI (LF-MMI)~\cite{Povey2016Purely} can be used to further improve the performance of hybrid ASR systems.

More recently there has also been a significant trend in the speech technology field moving from hybrid HMM-DNN system architecture to an all neural end-to-end (E2E) modeling paradigm. E2E systems represented by connectionist temporal classification (CTC)~\cite{graves2006connectionist}, RNN transducers (RNN-T)~\cite{rao2017exploring}, listen, attend and spell (LAS)~\cite{chan2016listen} or encoder-decoder with attention (AED)~\cite{bahdanau2016end}, LF-MMI~\cite{hadian2018end} and transformers~\cite{karita2019comparative}, utilize a single DNN model transforming an input acoustic feature sequence directly to a target stream of output tokens based on characters, word pieces or words.


A key task facing all ASR systems is to learn the systematic and latent variation among diverse speech data. This often creates a large mismatch between the training and evaluation data leading to recognition performance degradation. A major source of such variability is attributable to speaker level characteristics representing factors such as accent and idiosyncrasy, or physiological differences manifested in, for example, age or gender. To this end, speaker adaptation techniques play a vital role in current speech recognition systems. Speaker adaptation techniques used in current neural network based ASR systems can be characterized into several broad categories: auxiliary speaker embedding based approaches, feature transformation based methods, and model-based adaptation techniques. The model-based adaptation approaches exploit specially designed speaker-dependent (SD) DNN model parameters to compensate speaker level variability, for examples, the learning hidden unit contributions (LHUC)~\cite{swietojanski2014learning,swietojanski2016learning,zhang2016dnn} and parameterized activation functions (PAct)~\cite{siniscalchi2013hermitian,zhang2015parameterised,zhang2016dnn} techniques. More speaker adaptation approaches are introduced in Section \ref{sec:review}.

Model-based speaker adaptation approaches often require sufficient amounts of target speaker data for robust SD parameter estimation. However, the available amount of speaker-specific data is often limited in practical applications. Even when using compact SD parameter representation based on the aforementioned structured transforms, for example, an LHUC scaling vector of several thousand dimensions, speaker adaptation can still be prone to overfitting and poor generalization performance.  This issue is particularly prominent when rapid adaptation to individual speakers' voice characteristics using very little data, for example, a few seconds of speech, is required at the onset of user enrollment to ASR systems.

A solution proposed in this paper to address the inherent SD parameter uncertainty resulted from limited adaptation data is using Bayesian learning approaches. In the machine learning community, Bayesian learning has been established as a mathematically well formulated framework to account for modelling uncertainty in neural network systems including current deep learning models. A practical Bayesian framework to account for parameter uncertainty using Laplace approximation in back-propagation neural networks was first introduced in~\cite{Mackay1992A}. More efficient inference methods based on variational learning and Monte Carlo parameter sampling were proposed for Bayesian neural networks in~\cite{barber1998ensemble}, before being further extended to auto-encoder neural networks in~\cite{kingma2014stochastic}. In contrast, very limited previous research on Bayesian deep learning approaches has been conducted in the area of speech recognition. In~\cite{graves2011practical}, variational inference was adopted in the training of a Bayesian recurrent neural network (RNN) based phone recognition system using the TIMIT speech corpus. Bayesian learning for RNN language modeling was investigated in~\cite{Chien2016Bayesian}. More recently, a Bayesian learning framework was used to account for model uncertainty in sequence discriminative training of factored TDNN acoustic models~\cite{hu2019bayesian,hu2019lf}.
Stochastic noise injection to model parameters~\cite{murray1994enhanced} was also exploited to improve the generalization performance of E2E ASR systems~\cite{braun2019parameter,tuske2020single}.
The above application of Bayesian learning to current ASR systems focused on modelling parameter uncertainty in speaker independent systems, while the often more prominent data sparsity problem encountered during speaker adaptation on very limited data remains unvisited.


In order to address this issue, a full Bayesian learning based DNN speaker adaptation framework is proposed in this paper to account for SD parameter uncertainty given limited speaker specific adaptation data. The generic nature and advantages of this full Bayesian DNN adaptation framework is demonstrated via its application to three exemplar types of structured DNN speaker adaptation transforms: Bayesian learning of hidden unit contributions (BLHUC)~\cite{xie2019blhuc}, Bayesian parameterized activation functions (BPAct) and Bayesian hidden unit bias vectors (BHUB). In these Bayesian adaptation methods, deterministic SD parameters are replaced by latent variable posterior distributions to be learned for each speaker, thus allowing the parameter uncertainty to be explicitly modeled to reduce the risk of overfitting. The parameters of these posterior distributions are efficiently estimated using variational inference based approach further benefited from the use of Monte Carlo parameter sampling and re-parameterization. The proposed Bayesian DNN adaptation methods are investigated for both unsupervised test time speaker adaptation and speaker adaptive training.

Experiments conducted on state-of-the-art 300-hour speed perturbed Switchboard corpus trained LF-MMI factored TDNN and CNN-TDNN systems featuring i-vector speaker adaptation suggest the proposed Bayesian adaptation approaches consistently outperform the adapted system constructed using deterministic parameter based LHUC, parameterized activation functions (PAct) or hidden unit bias vectors (HUB) on the \emph{NIST Hub5'00} and \emph{RT03} evaluation sets in unsupervised test time speaker adaptation and speaker adaptive training. In particular, when using only the first five utterance from each speaker as the adaptation data, statistically significant absolute (relative) reduction up to 1.4\% (7.2\%) in word error rate (WER) were obtained on the \emph{CallHome} subset. Similar consistent performance improvements were also retained after LSTM recurrent neural network language model rescoring. The efficacy of the proposed Bayesian adaptation techniques is further demonstrated in a comparison against the state-of-the-art performance obtained on the same task using most recent hybrid and E2E systems reported in the literature.
The main contributions of this paper are summarized below:

1) To the best of our knowledge, this paper presents the first use of a full Bayesian framework to account for model uncertainty in speaker adaptation of DNN acoustic model for speech recognition. In contrast, previous research on Bayesian neural network techniques were restricted to the estimation of un-adapted acoustic models or language models in hybrid
systems~\cite{graves2011practical,Chien2016Bayesian,hu2019bayesian,hu2019lf} or E2E systems~\cite{braun2019parameter,tuske2020single}.

2) The efficient variational inference approach developed for the proposed Bayesian DNN adaptation uses a very few number of parameter samples (as low as one) to ensure their computational cost comparable to that of the baseline adaptation techniques. Their generic nature allows them to be applied to E2E approaches to address similar model uncertainty issues during adaptation to new speakers using limited data.

3) Statistically significant performance improvements over multiple model-based DNN adaptation methods given varying amounts of adaptation data were obtained on state-of-the-art baseline LF-MMI factored TDNN systems using i-vector feature based speaker adaptation~\cite{saon2013speaker,senior2014improving}. These demonstrate the power of the proposed Bayesian adaptation approaches to facilitate both rapid, instantaneous adaptation to individual speakers' voice, and batch mode adaptation performed using more data. Such flexibility is crucial for speech based conversational agents in commercial applications to ensure user satisfaction from their enrollment to the system very early on.

The rest of this paper is organized as follows. Section \ref{sec:review} introduces the related works of speaker adaptation. Section \ref{sec:adapt} reviews three types of structured DNN adaptation based on LHUC, hidden unit bias vectors and parameterized activation functions. A full Bayesian learning framework and its application to these three types of adaptation methods is proposed in Section \ref{sec:bl}. Section \ref{sec:exp} presents the experimental results and analysis of the Bayesian adaptation approaches conducted on state-of-the-art sequence discriminatively trained TDNN and CNN-TDNN acoustic models on a 300-hour \emph{Switchboard} setup. The last section drawns the conclusion and discusses possible future works.

\subsection{Related Works of Speaker Adaptation}
\label{sec:review}


In the auxiliary speaker embedding based approaches, auxiliary speaker embedding based DNN adaptation techniques encode speaker-dependent (SD) characteristics in a compact vector representation. The resulting SD embedding vectors are used as auxiliary input features to be fed into DNNs and facilitate speaker adaptation during both system training and evaluation. The SD auxiliary feature estimation is performed independently of the remaining recognition system components. For example, i-vectors~\cite{saon2013speaker,senior2014improving} are learned from Gaussian mixture model (GMM) based universal background models (UBMs).

In the feature transformation based methods, feature transforms can be applied to acoustic front-ends and aim to produce canonical, speaker invariant input features. These are then fed into the back-end DNN based recognition systems to model other remaining sources of variability, for example, longer span phonetic and phonological context dependency in speech. Feature-space maximum likelihood linear regression (f-MLLR) transforms~\cite{seide2011feature} are estimated at speaker level from GMM-HMM based ASR systems first~\cite{digalakis1995speaker,gales1998maximum} before being applied to acoustic front-ends to produce canonical, speaker invariant like input features that are then fed into DNN based recognition systems. In order to compensate for vocal tract length differences between speakers, physiologically motivated feature space adaptive transformations based on vocal tract length normalization (VTLN) can also be used~\cite{lee1996speaker,uebel1999investigation}. Speaker level VTLN normalized features can be obtained using either piecewise linear frequency warping factors directly applied to the spectrum, or affine linear transformations similar to f-MLLR.

In the model-based adaptation techniques, the amount of speaker specific adaptation data practically determines the SD parameters' modelling granularity. In order to ensure good generalization performance and mitigate the risk of overfitting, a trade-off between moving from adapting only parts of the whole acoustic model towards complete speaker dependent systems~\cite{liao2013speaker,yu2013kl} needs to be adjusted. To this end, existing researches focused on deriving compact forms of SD parameter representations, predominantly based on structured transforms that are positioned in various parts of the DNN acoustic model. These include the use of SD linear input networks (LIN)~\cite{neto1995speaker,li2010comparison}, linear output networks (LON)~\cite{li2010comparison}, linear hidden networks (LHN)~\cite{gemello2007linear}, learning hidden unit contributions (LHUC)~\cite{swietojanski2014learning,swietojanski2016learning,zhang2016dnn}, parameterized activation functions (PAct)~\cite{zhang2015parameterised,zhang2016dnn}, subspace factorized speaker-independent (SI) and SD affine transformations~\cite{zhao2016low}, speaker code based bias vectors~\cite{xue2014fast}, and adaptive linear interpolation of basis sub-networks' hidden outputs~\cite{wu2015multi,tan2016cluster}, akin to cluster based speaker adaptation for GMM-HMM systems~\cite{gales2000cluster}. In addition to only modelling the variability among speakers found in the test data during recognition, the estimation of SD parameters in both the system training and evaluation stages leads to speaker adaptive training (SAT)~\cite{anastasakos1996compact}, thus allowing a joint optimization of both the SD and SI parameters shared among all speakers in the training data.


Other forms of model-based approaches learn to compensate speaker level variability without explicitly using SD parameters. For example, adversarial learning~\cite{meng2018speaker,tsuchiya2018speaker} of DNN acoustic models can be used to generate speaker-invariant hidden layer outputs. Although most of the above adaptation approaches were first developed for hybrid ASR systems, recent researches suggest these can be further applied to E2E based ASR systems~\cite{li2018speaker,meng2019speaker,sim2019investigation}.


A series of speaker adaptation techniques have been developed in the past to address the data sparsity issue. Among these, data augmentation methods were used to expand the target speaker or domain data using, for example, perturbation of speaking rates~\cite{ko2015audio,huang2020acoustic} or vocal tract length~\cite{jaitly2013vocal}, speaker mappings~\cite{fainberg2016improving} and synthesis techniques~\cite{huang2020using}. Another category of solutions to this problem is based on regularized adaptation methods. For example, L2 regularization was used to prevent the SD parameters from diverging too far from those of the SI models~\cite{liao2013speaker}. This is similar to maximum a posteriori (MAP) adaptation~\cite{huang2015maximum,huang2016unified,huang2016bayesian,huang2017hierarchical} which restricts the estimation of SD parameters using a prior distribution empirically estimated, for example, over a set of SD adapted parameters obtained from the training data. Furthermore, Kullback-Leibler (KL) divergence between the output distributions of SI model and adapted model was also exploited for regularization of DNN adaptation~\cite{yu2013kl}. Hyper-parameter grid-search and meta-learning based adaptation methods regularized by early stopping and appropriate learning rate settings have also been investigated~\cite{klejch2018learning}.

\section{Model-Based Speaker Adaptation}
\label{sec:adapt}

In this section, we introduce several forms of widely used DNN model-based speaker adaptation approaches. A common idea of applying model-based speaker adaptation to a DNN model is to modify a set of SD parameters for each speaker.
A widely used form of SD parameters is an affine transformation deployed over hidden vectors in a DNN layer.
However, the use of full matrix affine linear transformation with thousands of parameters makes the adaptation prone to overfitting. An alternative idea is to use less SD parameters for adaptation, for example, in form of vectors with significantly lower dimension than affine transformations.

\subsection{Learning Hidden Unit Contributions}
\label{sec:lhuc}

\begin{figure*}[!htbp]
  \begin{center}
    \includegraphics[width=17cm]{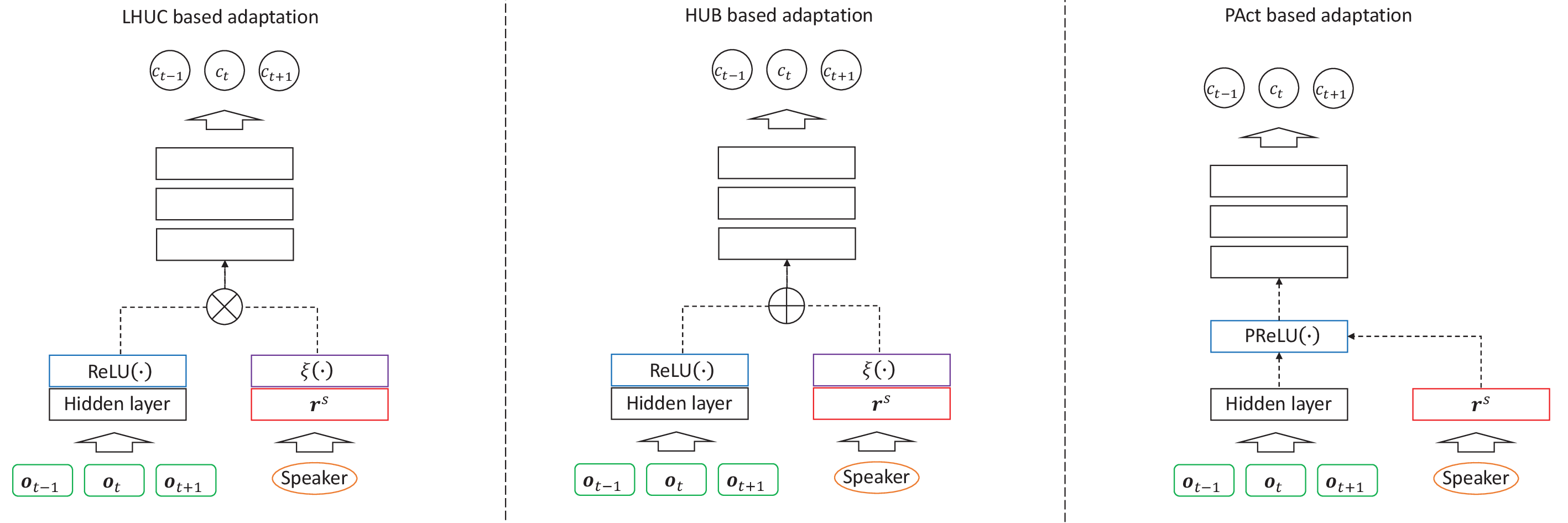}
    \caption{Examples of applying learning hidden unit contributions (LHUC) adaptation (left), hidden unit bias vectors (HUB) adaptation (middle), and parameterized activation functions (PAct) adaptation (right) to a hidden layer of DNN acoustic model.}
    \label{fig:adapt_ml}
  \end{center}
\end{figure*}

An intuitive solution is to reduce the number of SD parameters by restricting the affine transformation as a diagonal matrix. The SD parameters can then be realized as an SD scaling vector to the DNN hidden layer vector. This approach is known as the learning hidden unit contributions (LHUC) adaptation~\cite{swietojanski2016learning}. In a certain DNN layer and for a certain speaker $s$, letting ${\bm h}_t \in \mathbb{R}^D$ be a hidden layer vector at the $t$th instant, and ${\bm r}^{s}$ be the SD parameters for speaker $s$, the adapted hidden vector is computed as
\begin{eqnarray}
{\bm h}^{s}_t 
 & = & \xi({\bm r}^{s}) \otimes {\bm h}_t
\label{eq:scale_offset}
\end{eqnarray}
where $\xi({\bm r}^{s})$ is the scaling vector parameterized by ${\bm r}^{s} \in \mathbb{R}^{D_{r}}$, and $\otimes$ denotes the Hadamard product operation. In practice, ${\bm h}_t$ can be the hidden vector computed after or prior to a nonlinear activation function. In this study, ${\bm h}_t$ is a ReLU activated hidden vector. The LHUC adaptation is shown as the left part in figure \ref{fig:adapt_ml}.

The dimension $D_{r}$ is commonly not larger than the layer dimension $D$. The LHUC activation function $\xi: \mathbb{R}^{D_{r}} \rightarrow \mathbb{R}^D$ constraints the ranges of the LHUC scaling. If $D_{r} = D$, a widely used form of LHUC activation is $2\text{Sigmoid}$ such that the scaling range is $(0,2)$~\cite{swietojanski2014learning}. Other functions can be ReLU~\cite{zhang2016dnn}, exponential~\cite{swietojanski2016learning}, or identity functions. If $D_{r}$ is less than $D$, $\xi$ can be a linear projection matrix mapping lower-dimensional SD parameters to high-dimensional scaling vectors~\cite{samarakoon2016subspace}.

\subsection{Hidden Unit Bias Vector}
\label{sec:ofs}

Another way to reduce the number of SD parameters in affine transformation is to use the bias vector only. The SD parameters serve as a bias or offset to the hidden vector in a certain DNN layer. This is referred to hidden unit bias vectors (HUB) adaptation in this study. The adapted hidden vector is then computed as
\begin{eqnarray}
{\bm h}^{s}_t 
 & = & {\bm h}_t + \xi({\bm r}^{s})
\label{eq:scale_offset}
\end{eqnarray}
where $\xi({\bm r}^{s})$ is the bias vector parameterized by ${\bm r}^{s} \in \mathbb{R}^{D_{r}}$. The speaker code based adaptation~\cite{xue2014fast} is one of the approaches using only the bias vector as SD parameters. The HUB adaptation is shown as the middle part in figure \ref{fig:adapt_ml}.
Similar to LHUC adaptation, the bias activation function $\xi$ can be function of Tanh, identity, or a linear projection matrix mapping lower-dimensional SD parameters to high-dimensional bias vectors~\cite{xue2014fast}.

\subsection{Parameterised Activation Functions}
\label{sec:prelu}

The third type of structured DNN transforms based adaptation considered in this paper uses parameterised activation functions (PAct)~\cite{zhang2015parameterised}. When using parameterised ReLU (PReLU) as activation function, and defining the SD piece-wise linear slope parameters as ${\bm r}^{s} = [ {\bm \alpha}^{s}, {\bm \beta}^{s} ]^T \in \mathbb{R}^{2D}$, the $d$th element of speaker-specific activation vector after adaptation is computed as

\begin{equation}
\label{eq:prelu}
h^{s}_{t,d} = \left\{
\begin{aligned}
{\alpha}^{s}_d z_{t,d} & , & z_{t,d} > 0 \\
{\beta}^{s}_d z_{t,d} & , & z_{t,d} \leq 0
\end{aligned}
\right.
\end{equation}

where ${\alpha}^{s}_d$ and ${\beta}^{s}_d$ are the $d$th elements of ${\bm \alpha}^{s}$ and ${\bm \beta}^{s}$ respectively, and $z_{t,d}$ is the hidden node value prior to the PReLU activation. The PAct adaptation is shown as the right part in figure \ref{fig:adapt_ml}.
The standard ReLU activation can be viewed as a PReLU activation with ${\bm \alpha}^{s} = 1$ and ${\bm \beta}^{s} = 0$.

\subsection{Estimation with Minimum Cross Entropy}
\label{sec:ML}

The LHUC, HUB, and PAct adaptation approaches share the same framework for SD parameter estimation. The SD parameters can be estimated by minimizing cross entropy (CE). The minimum CE estimator can be given by
\begin{eqnarray}
\hat{{\bm r}}^{s}_{\text{CE}} & = & \arg\min_{{\bm r}^{s}} \{ - \log P(C^{s}|{\bm O}^{s},{\bm r}^{s}) \}
\label{eq:ml}
\end{eqnarray}
where ${\bm O}^{s}$ denotes the adaptation data set for speaker $s$, and $C^{s}$ is the corresponding supervision labels, e.g., HMM state alignments.
Without loss of generality, the SI parameters in the model is omitted.

In the context of unsupervised test-time adaptation, the adaptation data ${\bm O}^{s}$ is commonly a part of test data, hence the supervision $C^{s}$ is not available during adaptation. In order to handle the problem, $C^{s}$ can be derived by applying decoding process to the adaptation data ${\bm O}^{s}$. After SD parameters estimated, if necessary, the adaptation data can be decoded again to re-generate the supervision $C^{s}$ for re-estimation of the SD parameters. Finally, the estimated SD parameters are employed by decoding process.

In the speaker adaptive training (SAT) framework~\cite{anastasakos1996compact}, SD parameters associated with speakers in training data is estimated in the training stage. This enables the SI parameters to be learned more effectively with less distortion by speaker variability. In this stage, the SD parameters are updated with the same loss function as that for DNN model training.

In the decoding process, given test data $\tilde{{\bm O}}^s$, the corresponding states $\tilde{C}^s$ is inferred by the adapted model with probability
\begin{eqnarray}
\!\!\!\!\! P(\tilde{C}^s | \tilde{{\bm O}}^s, {\bm O}^{s}, C^{s})
& \approx & P(\tilde{C}^s | \tilde{{\bm O}}^s, \hat{{\bm r}}^{s}_{\text{CE}})
\label{eq:ml_infer}
\end{eqnarray}
where $\hat{{\bm r}}^{s}_{\text{CE}}$ is the minimum CE estimate of ${\bm r}^{s}$ in equation (\ref{eq:ml}).

\section{Bayesian Learning of Adaptation Approach}
\label{sec:bl}

This section presents a framework for applying full Bayesian learning to DNN model-based speaker adaptation approaches. This framework is demonstrated via applications to the aforementioned three DNN adaptation approaches: Bayesian learning of hidden unit contributions (BLHUC), Bayesian hidden unit bias vectors (BHUB), and Bayesian parameterized activation functions (BPAct).

\subsection{Inference of Bayesian Learning}
\label{sec:infer}

In contrast to the conventional fixed parameter based decoding process determining the most likely word or HMM state sequence, the comparable Bayesian inference incorporating the adaptation parameter uncertainty is given as the following
\begin{eqnarray}
\!\!\!\!\! P(\tilde{C}^s | \tilde{{\bm O}}^s, {\bm O}^{s}, C^{s}) \!\!\!\! & = & \!\!\!\! \int P(\tilde{C}^s | \tilde{{\bm O}}^s,{\bm r}^{s}) p({\bm r}^{s}|{\bm O}^{s}, C^{s}) d{\bm r}^{s}. \label{eq:bayes_infer}
\end{eqnarray}
The inference processes of minimum CE and MAP~\cite{huang2015maximum} are special cases of the Bayesian inference. When using the minimum CE estimate for inference as equation (\ref{eq:ml_infer}), an approximation is employed by assuming that $P(\hat{{\bm r}}^{s}_{\text{CE}}|{\bm O}^{s}, C^{s}) \approx 1$. Similarly, the MAP adaptation makes use of $\hat{{\bm r}}^{s}_{\text{MAP}} = \arg\max_{{\bm r}^{s}} \log P({\bm r}^{s}|{\bm O}^{s}, C^{s})$ by assuming $P(\hat{{\bm r}}^{s}_{\text{MAP}}|{\bm O}^{s}, C^{s}) \approx 1$.
Under these assumptions,
the point estimates may introduce estimation bias to the inference process.

In Bayesian learning, the deterministic SD parameters are replaced by probabilistic latent variables. The same notation ${\bm r}^{s}$ is used for the latent variable without loss of generality.
Assuming that the posterior distribution $p({\bm r}^{s} | {\bm O}^{s}, C^{s})$ is given, the equation (\ref{eq:bayes_infer}) can be directly used for decoding. The Monte Carlo approximation of the integral in equation (\ref{eq:bayes_infer}) with $J_{\text{inf}}$ samples is computed as
\begin{eqnarray}
P(\tilde{C}^s | \tilde{{\bm O}}^s, {\bm O}^{s}, C^{s}) & \approx & \frac {1} {J_{\text{inf}}} \sum_{j=1}^{J_{\text{inf}}} P(\tilde{C}^s | \tilde{{\bm O}}^s,{\bm r}^{s}_j)
\label{eq:bayes_infer_mc}
\end{eqnarray}
where ${\bm r}^{s}_j$ denotes the $j$th random sampling of ${\bm r}^{s}$ drawn from the posterior distribution $p({\bm r}^{s} | {\bm O}^{s}, C^{s})$.

However, computing the equation (\ref{eq:bayes_infer_mc}) over $J_{\text{inf}}$ random samples could be computationally more expensive than using minimum CE point estimate.
The expectation of ${\bm r}^{s}$ is an alternative point estimate to approximate the integral for inference. This is given by
\begin{eqnarray}
P(\tilde{C}^s | \tilde{{\bm O}}^s, {\bm O}^{s}, C^{s}) & \approx & P(\tilde{C}^s | \tilde{{\bm O}}^s, \mathbb{E}[{\bm r}^{s}|{\bm O}^{s}, C^{s}])
\label{eq:bayes_infer_exp}
\end{eqnarray}
where $\mathbb{E}[\cdot]$ denotes the expectation.
Using the expectation takes a similar complexity and procedure to using the minimum CE estimate by assuming that $P(\mathbb{E}[{\bm r}^{s}|{\bm O}^{s}, C^{s}]|{\bm O}^{s}, C^{s}) \approx 1$. Moreover, it normally represents an unbiased average over the possible latent variable values.
In order to use Bayesian learning for adaptation, the key issue is to compute the posterior distribution $p({\bm r}^{s}|{\bm O}^{s}, C^{s})$ over the SD latent variable ${\bm r}^{s}$.

\subsection{Framework for Variational Approximation}
\label{sec:var}

Variational distribution~\cite{mackay2003information} $q_s({\bm r}^{s})$ can be used to approximate the true posterior distribution $p({\bm r}^{s}|{\bm O}^{s},C^{s})$.
For the speaker $s$, the marginal cross entropy over the adaptation data set $\{C^{s}, {\bm O}^{s}\}$ can be computed as
\begin{eqnarray}
&& - \log P(C^{s}|{\bm O}^{s}) = - \log \int P(C^{s},{\bm r}^{s}|{\bm O}^{s}) d{\bm r}^{s} \nonumber \\
&& \leq -  \underbrace{\int q_s({\bm r}^{s}) \log P(C^{s}|{\bm O}^{s}, {\bm r}^{s}) d{\bm r}^{s}}_{\mathcal{L}_1(q_s)} +  KL(q_s||p_0) \nonumber \\
&& \overset{\text{def}}{=} \mathcal{L}(q_s)
\label{eq:vb}
\end{eqnarray}
where $KL(q_s||p_0) = \int q_s({\bm r}^{s}) \log \frac{q_s({\bm r}^{s})}{p_0({\bm r}^{s})} d{\bm r}^{s}$ denotes the KL divergence between the variational distribution $q_s({\bm r}^{s})$ and prior distribution $p_0({\bm r}^{s})$ of ${\bm r}^{s}$. The gap between the cross entropy $- \log P(C^{s}|{\bm O}^{s})$ and its upper bound $\mathcal{L}(q_s)$ is actually the KL divergence between the true posterior $p({\bm r}^{s}|{\bm O}^{s},C^{s})$ and the variational distribution $q_s({\bm r}^{s})$, which is expressed as
\begin{eqnarray}
KL(q_s({\bm r}^{s})||p({\bm r}^{s}|{\bm O}^{s},C^{s})) \!\!\!\!\! & = & \!\!\!\!\! \int q_s({\bm r}^{s}) \log \frac{q_s({\bm r}^{s})}{p({\bm r}^{s}|{\bm O}^{s},C^{s})} d{\bm r}^{s} \nonumber \\
\!\!\!\!\! = && \!\!\!\!\!\!\!\!\!\!\!\!\!\!\! \mathcal{L}(q_s) - \left\{ - \log P(C^{s}|{\bm O}^{s}) \right\}.
\end{eqnarray}
Obviously, minimizing the upper bound $\mathcal{L}(q_s)$ is equivalent to minimizing the distance between $p({\bm r}^{s}|{\bm O}^{s},C^{s})$ and $q_s({\bm r}^{s})$. Thereby, the goal is to estimate $q_s$ for minimizing $\mathcal{L}(q_s)$.

For simplicity, both $q_s$ and $p_0$ are assumed to be normal distributions parameterized by ${\bm \theta}^{\text{B}}_s = \{{\bm \mu}_s, {\bm \sigma}_s\}$ and ${\bm \theta}^{\text{B}}_0 = \{{\bm \mu}_0, {\bm \sigma}_0\}$ respectively. These distributions can be written as
\begin{eqnarray}
q_s(r^{s}_d) = \mathcal{N}(r^{s}_d; \mu_{s,d}, \sigma_{s,d}^2) \nonumber \\
p_0(r^{s}_d) = \mathcal{N}(r^{s}_d; \mu_{0,d}, \sigma_{0,d}^2)
\end{eqnarray}
where $\mu_{s,d}$ and $\mu_{0,d}$ denote the $d$th elements of the mean vectors ${\bm \mu}_s$ and ${\bm \mu}_0$, and $\sigma_{s,d}$ and $\sigma_{0,d}$ denote the $d$th elements of the standard deviation vectors ${\bm \sigma}_s$ and ${\bm \sigma}_0$, respectively.

By this assumption, the random sample ${\bm r}^{s}_j$ for inference in equation (\ref{eq:bayes_infer_mc}) is drawn from the distribution $\mathcal{N}({\bm \mu}_{s}, {\bm \sigma}_{s}^2)$, and the inference in equation (\ref{eq:bayes_infer_exp}) is written as
\begin{eqnarray}
P(\tilde{C}^s | \tilde{{\bm O}}^s, {\bm O}^{s}, C^{s}) & \approx & P(\tilde{C}^s | \tilde{{\bm O}}^s, {\bm \mu}_{s})
\label{eq:bayes_infer_mu}
\end{eqnarray}
where the expectation of posterior distribution is simply approximated by ${\bm \mu}_{s}$, which is presented as
\begin{eqnarray}
\mathbb{E}[{\bm r}^{s}|{\bm O}^{s}, C^{s}] \approx {\bm \mu}_{s}.
\end{eqnarray}
Meanwhile, the KL divergence $KL(q_s||p_0)$ has a closed form integral that is explicitly calculated as
\begin{eqnarray}
K\!L(q_s||p_0) \!=\! \frac{1}{2} \! \sum_{d=1}^D \!\left\{\!\! \frac{(\mu_{s,d} \!-\! \mu_{0,d})^2 + \sigma_{s,d}^2} {\sigma_{0,d}^2} \!-\! \log \frac{ \sigma_{s,d}^2}{\sigma_{0,d}^2} - 1 \! \right\}\!\!.\!\!
\label{eq:kl}
\end{eqnarray}

However, directly updating ${\bm \theta}^{\text{B}}_s$ in the integral of upper bound $\mathcal{L}(q_s)$ is nontrivial. The re-parameterization introduced in~\cite{kingma2014stochastic} is an efficient technique to obtain a differentiable estimator of the upper bound.
The parameters ${\bm \theta}^{\text{B}}_s$ can be re-parameterized as
\begin{eqnarray}
{\bm r}^{s} = {\bm \mu}_s + {\bm \sigma}_s \otimes {\bm \epsilon}
\label{eq:random}
\end{eqnarray}
where ${\bm \epsilon} \sim \mathcal{N}({\bm 0},{\bm I})$. Applying Monte Carlo approximation, the integral in equation (\ref{eq:vb}) is rewritten as
\begin{eqnarray}
\mathcal{L}_1(q_s) & = & \int q_s({\bm r}^{s}) \log P(C^{s}|{\bm O}^{s}, {\bm r}^{s}) d{\bm r}^{s} \nonumber \\
& = & \int \mathcal{N}({\bm \epsilon};0,I) \log P(C^{s}|{\bm O}^{s}, {\bm \mu}_s + {\bm \sigma}_s \otimes {\bm \epsilon}) d{\bm \epsilon} \nonumber \\
& \approx & \frac{1}{J_{\text{est}}} \sum_{j=1}^{J_{\text{est}}} \log P(C^{s}|{\bm O}^{s}, {\bm \theta}^{\text{B}}_s, {\bm \epsilon}_j)
\label{eq:integal}
\end{eqnarray}
where ${\bm \epsilon}_j$ is the $j$th random sampling of ${\bm \epsilon}$ drawn from the standard normal distribution $\mathcal{N}({\bm 0},{\bm I})$.

To minimize the upper bound $\mathcal{L}(q_s)$ in equations (\ref{eq:vb}) computed by equations (\ref{eq:kl}) and (\ref{eq:integal}), the gradient of ${\bm \theta}^{\text{B}}_s$ at the $m$th update for speaker $s$ is computed as
\begin{eqnarray}
&& \!\!\!\!\!\!\!\!\!\!\!\!\! \frac{\partial \mathcal{L}^{(m)}(q_s)}{\partial {\bm \theta}^{\text{B}}_s} \label{eq:BLHUC:update0} \\
&& \!\!\!\!\!\!\!\!\!\!\!\!\!  \approx \! -\frac{N_s}{N_{m,s}} \frac{1}{J_{\text{est}}} \!\! \sum_{j=1}^{J_{\text{est}}} \frac{\partial \log P(C^{s}_m|{\bm O}^{s}_m, {\bm \theta}^{\text{B}}_s, {\bm \epsilon}_j)}{\partial \theta^{\text{B}}_s} \!+\! \lambda \frac{\partial KL(q_s||p_0)}{\partial {\bm \theta}^{\text{B}}_s} \nonumber
\label{eq:BLHUC:update0}
\end{eqnarray}
where $N_s$ denotes the number of frames for speaker $s$ in all adaptation data, $N_{m,s}$ denotes the number of frames for speaker $s$ used by the $m$th update,
$\{C^{s}_m,{\bm O}^{s}_m\} \subseteq \{C^{s},{\bm O}^{s}\}$ is the adaptation data used in the $m$th update for speaker $s$,
and $\lambda \in (0,1]$ makes the gradient parts computed from the DNN and KL divergence have similar scales.
Specifically, the gradients of upper bound $\mathcal{L}^{(m)}(q_s)$ over parameters ${\bm \mu}_s$ and ${\bm \sigma}_s$ on the $d$th dimension can be calculated as
\begin{eqnarray}
&& \!\!\!\!\!\!\!\!\!\!\! \frac{\partial \mathcal{L}^{(m)}(q_s)}{\partial \sigma_{s,d}}  \!=\!  \frac{N_s}{N_{m,s}} \frac{1}{J_{\text{est}}}  \!\! \sum_{j=1}^{J_{\text{est}}} G_{m,s,j,d} \epsilon_{j,d}  +  \lambda \! \left( \! \frac{\sigma_{s,d}}{\sigma_{0,d}^2} \!-\! \frac{1}{\sigma_{s,d}} \! \right) \nonumber \\
&& \!\!\!\!\!\!\!\!\!\!\! \frac{\partial \mathcal{L}^{(m)}(q_s)}{\partial \mu_{s,d}}  \!=\!  \frac{N_s}{N_{m,s}}  \frac{1}{J_{\text{est}}} \!\! \sum_{j=1}^{J_{\text{est}}} G_{m,s,j,d}  +  \lambda  \frac{\mu_{s,d} - \mu_{0,d}}{\sigma_{0,d}^2}
\label{eq:BLHUC:update}
\end{eqnarray}
where $G_{m,s,j,d} = -\frac{\partial \log P(C^{s}_m|{\bm O}^{s}_m, {\bm r}^{s}_j)}{\partial r^{s}_{j,d}}$ is the gradient computed from DNN model.

The gradient $G_{m,s,j,d}$ in equation (\ref{eq:BLHUC:update}) has the same form as that derived by minimum CE estimation using parameters ${\bm r}^{s}_j = {\bm \mu}_s + {\bm \sigma}_s \otimes {\bm \epsilon}_j$, thus has different forms for different adaptation approaches:
\subsubsection{\bf Bayesian LHUC adaptation}
For using LHUC adaptation approaches as introduced in Section \ref{sec:lhuc}, the Bayesian learning based adaptation approach is referred to Bayesian LHUC (BLHUC) adaptation. The corresponding gradient $G_{m,s,j,d}$ is given by
\begin{eqnarray}
&& \!\!\!\!\!\!\!\!\!\!\!\!\!\!\!\!\!\!\!\!\! G_{m,s,j,d}^{\text{LHUC}} = - \! \sum_{t_m} \! \sum_{d'=1}^D \! \frac{\partial \log P(C^{s}_m|{\bm O}^{s}_m, {\bm r}^{s}_j)}{\partial h^{s}_{t_m,d'}} \frac{\partial \xi_{d'}({\bm r}^{s}_j)} {\partial r^{s}_{j,d}} h_{t_m,d'}
\end{eqnarray}
where $t_m$ denotes a time instant in the $m$th update. The BLHUC adaptation is shown as the left part in figure~\ref{fig:adapt_bl}.
\subsubsection{\bf Bayesian HUB adaptation}
For HUB adaptation as introduced in Section \ref{sec:ofs}, the Bayesian learning approach is referred to Bayesian HUB (BHUB) adaptation. The gradient $G_{m,s,j,d}$ is given by
\begin{eqnarray}
&& \!\!\!\!\!\!\!\!\!\!\!\!\!\!\!\! G_{m,s,j,d}^{\text{HUB}} = - \! \sum_{t_m} \! \sum_{d'=1}^D \! \frac{\partial \log P(C^{s}_m|{\bm O}^{s}_m, {\bm r}^{s}_j)}{\partial h^{s}_{t_m,d'}} \frac{\partial \xi_{d'}({\bm r}^{s}_j)} {\partial r^{s}_{j,d}}.
\end{eqnarray}
The BHUB adaptation is shown as the middle part in figure~\ref{fig:adapt_bl}.
\subsubsection{\bf Bayesian PAct adaptation}
\begin{figure*}[!htbp]
  \begin{center}
    \includegraphics[width=17cm]{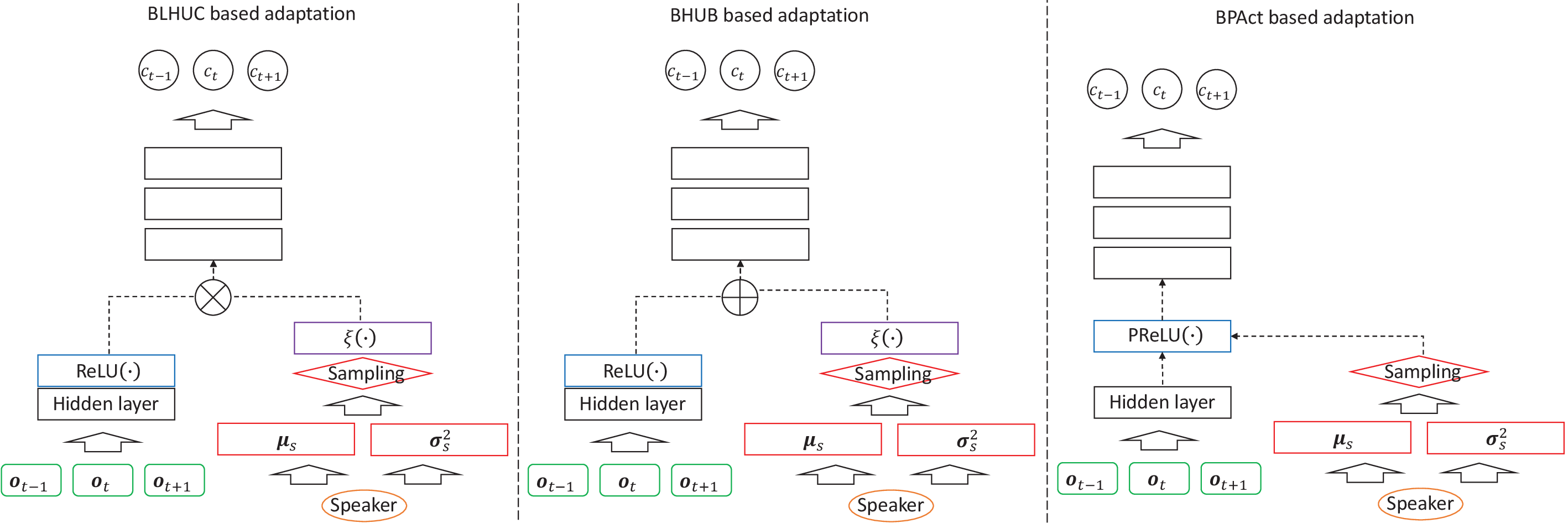}
    \caption{Examples of applying Bayesian LHUC (BLHUC) adaptation (left), Bayesian hidden unit bias vectors (BHUB) adaptation (middle), and Bayesian parameterized activation functions (BPAct) adaptation (right) to a hidden layer of DNN acoustic model.}
    \label{fig:adapt_bl}
  \end{center}
\end{figure*}

For PAct adaptation as introduced in Section \ref{sec:prelu}, the gradient $G_{m,s,j,d}$ becomes
\begin{eqnarray}
&& \!\!\!\!\!\!\!\!\!\!\!\!\!\!\!\! G_{m,s,j,d}^{\text{PAct}_{\alpha}} = - \! \sum_{t_m} \! \frac{\partial \log P(C^{s}_m|{\bm O}^{s}_m, {\bm r}^{s}_j)}{\partial \{ {\alpha}^{s}_{j,d} z_{t_m,d} \} } z_{t_m,d}
\end{eqnarray}
\begin{eqnarray}
&& \!\!\!\!\!\!\!\!\!\!\!\!\!\!\!\! G_{m,s,j,d}^{\text{PAct}_{\beta}} = - \! \sum_{t_m} \! \frac{\partial \log P(C^{s}_m|{\bm O}^{s}_m, {\bm r}^{s}_j)}{\partial \{ {\beta}^{s}_{j,d} z_{t_m,d} \} } z_{t_m,d}
\end{eqnarray}
with ${\bm r}^{s} = [{\bm \alpha}^{s}, {\bm \beta}^{s}]^T$. This Bayesian learning approaches is referred to Bayesian PAct (BPAct) adaptation, and is shown as the right part in figure~\ref{fig:adapt_bl}.

\subsection{Implementation Details}
\label{sec:implement}

\subsubsection{\bf Choices of Activation Functions}
\label{sec:act}

The activation function $\xi$ in Bayesian adaptation affects its performance. Different forms of $\xi$ can be used in BLHUC and BHUB adaptation as mentioned in Sections \ref{sec:lhuc} and \ref{sec:ofs}. In this paper we assume $D_{r} = D$ (e.g., the same dimensionality for both the hidden vector and LHUC SD parameters in one layer), and the forms of $\xi$ for BLHUC and BHUB adaptation are given in Table \ref{tab:act}.
\begin{table}[!htbp]
  \begin{center}
  {
    \caption{The choices of activation functions and their corresponding prior distributions for BLHUC, BHUB, and BPAct adaptation. }
    \label{tab:act}
    \scalebox{0.9}[0.9]
    {
    \begin{tabular}{l|c|l|l} \toprule[0.7pt]
          Bayesian adaptation & Latent variable & Activation $\xi$ & Prior $p_0$ \\ \midrule[0.7pt]
          \multirow{3}{*}{BLHUC} & \multirow{3}{*}{${\bm r}^{s}$} & Identity & $\mathcal{N}(1,1)$ \\
                               &  & 2Sigmoid & $\mathcal{N}(0,1)$ \\
                               &  & Exponential    & $\mathcal{N}(0,1)$ \\ \hline
          \multirow{2}{*}{BHUB} & \multirow{2}{*}{${\bm r}^{s}$} & Identity & $\mathcal{N}(0,0.01)$ \\
                                &  & Tanh & $\mathcal{N}(0,0.01)$ \\ \hline
          \multirow{2}{*}{BPAct} & ${\bm \alpha}^{s}$ & - & $\mathcal{N}(1,1)$ \\
                                & ${\bm \beta}^{s}$ & - & $\mathcal{N}(0,1)$ \\ \bottomrule[0.7pt]
    \end{tabular}
    }
  }
  \end{center}
\end{table}

\subsubsection{\bf Choices of Prior Distribution}
\label{sec:prior}

One important issue of applying Bayesian adaptation is the choice of prior distribution, which is dependent on the form of activation function mentioned in Section \ref{sec:act}.
The prior distribution $p_0$ can be directly given as a certain simple distribution, which is shown as in Table \ref{tab:act}.
Alternatively, an empirical prior distribution can be used similarly to that in MAP adaptation~\cite{huang2015maximum}. The empirical prior is estimated using training data by applying minimum CE estimation to the SD parameters.
Letting ${\bm r}^{s}$ be the SD parameter estimates to training data, the $d$th element of the mean and variance vectors of the prior distribution are computed over all training speakers as
\begin{eqnarray}
\mu_{0,d} & = & \frac{1}{S} \sum_{s=1}^S r^{s}_d \nonumber \\
\sigma_{0,d}^2  & = & \frac{1}{S} \sum_{s=1}^S (r^{s}_d - \mu_{0,d})^2
\label{eq:prior_estimate}
\end{eqnarray}
where $S$ is the number of speakers in the training data.

\subsubsection{\bf Test-Time Adaptation and SAT}
\label{sec:test}

During test-time adaptation using Bayesian learning, SD latent variable distributions are learned using variational inference and applied to the ReLU activated hidden vectors for different layers. All the other parameters in the baseline SI system shared among all speakers are fixed. Cross entropy is minimized by back-propagation based on canonical gradients in 7 epochs. The HMM state alignments used as supervision labels for adaptation are obtained by decoding using the baseline SI system. In the first $n$ layer applying Bayesian adaptation, the batch level gradient scaling $\lambda$ in equation (\ref{eq:BLHUC:update0}) is set empirically to $\min\{10^{n-5},1\}$. In the SAT framework, the adaptation approaches using deterministic SD parameters are applied in the training stage, because data for training speakers should be sufficient such that Bayesian learning is unnecessary.



\subsubsection{\bf Computational Efficiency in Inference}
\label{sec:cost}

The variational inference based Bayesian adaptation approaches presented in Section \ref{sec:infer} requires Monte Carlo sampling in order to approximate integral in equations (\ref{eq:bayes_infer}) and (\ref{eq:bayes_infer_mc}) for inference, and to approximate the first term of the marginal log-likelihood shown in equations (\ref{eq:vb}) and (\ref{eq:integal}). The number of samples drawn has a direct impact on the computational efficiency during inference. In order to maintain the overall computational cost of Bayesian adaptation to that of the conventional deterministic parameter adaptation methods, the following approaches can be used.
First, as mentioned in Section \ref{sec:infer}, taking the parameter mean or expectation during inference as equation (\ref{eq:bayes_infer_exp}) is more efficient than drawn samples using equation (\ref{eq:bayes_infer_mc}).
Second, it is feasible to use only one sample in each update~\cite{kingma2014stochastic} of Bayesian adaptation in equation (\ref{eq:BLHUC:update0}), i.e., $J_{\text{est}} = 1$.
Finally, in each layer the Gaussian variances $\sigma_{s,d}^2$ of SD latent variable in a posterior distribution can be tied together over each of the dimension $d$.
This allows for each speaker, when using any of three adaptation methods (LHUC, HUB or PAct), the effective number of SD parameters of their latent posterior distributions are comparable to that of using conventional deterministic parameter adaptation.
All these techniques allow Bayesian adaptation to have comparable computational efficiency to the deterministic adaptation.


\section{Experiments}
\label{sec:exp}

In this experiment section, the performance of the Bayesian learning based speaker adaptation approaches including BLHUC, BHUB, and BPAct proposed in Section III are investigated on the commonly used 300-hour \emph{Switchboard} conversational telephone speech dataset and its augmented 900-hour version derived using speed perturbation~\cite{ko2015audio}. This section is organized as follows. Section \ref{sec:data} introduces the \emph{Switchboard} speech data used in the experiments.  A series of experiments conducted on the 300-hour setup are then presented in Section \ref{sec:300hr}. The baseline TDNN system trained on the 300-hour data is described in Section \ref{sec:baseline}. An initial set of experiments in Section \ref{sec:bayes} serve to confirm the implementation details and settings previously given in Section \ref{sec:implement}, which are sensible to use for the subsequent experiments in the rest of this paper. The robustness of our Bayesian adaptation methods is then demonstrated in Section \ref{sec:amount} by performing adaptation using varying amounts of speaker level data, as little as five utterances of speech.  The advantages of the Bayesian adaptation techniques of this paper are further shown in Section \ref{sec:other} through an extensive comparison drawn against existing regularised adaptation methods. In Section \ref{sec:lstm} the performance of Bayesian adaptation methods is further evaluated on speed perturbed 900-hour setup featuring more advanced baseline system configurations using i-vector adapted CNN-TDNNs and LSTM-RNN language models. A final performance contrast against a set of latest hybrid and end-to-end systems reported in the literature is presented in Section \ref{sec:adv} to further demonstrate the superiority of Bayesian adaptation approaches.


\subsection{Data Description}
\label{sec:data}

Conversational telephone speech is a widely investigated ASR task in which speaker variability is considered a major challenge. The Switchboard-1 Release 2 (LDC97S62) corpus~\cite{godfrey1992switchboard} was used in the training stage. Near 286 hours of speech collected from 4804 speakers were utilized as training set. Approximate 3.6 minutes of speech data was recorded from each speaker on average. This training set was divided into 191,602 utterances of which each lasted for about 5.4 seconds on average.

For evaluation, the results were reported on the \emph{NIST Hub5'00} (LDC2002S09) and \emph{RT03} (LDC2007S10) data sets.
The \emph{Hub5'00} set consisted of two subsets with different domain properties: \emph{Switchboard (SWBD)} and \emph{CallHome (CHE)}. The \emph{SWBD} subset with 2 hours of speech was better matched to the training data. The \emph{CHE} subset with 1.5 hours of speech had larger mismatch thus was a more challenging task. Both of these two subsets were collected from 40 speakers. Approximate 3.1 and 2.5 minutes of speech data were recorded from each speaker in the \emph{SWBD} and \emph{CHE} subsets respectively on average. Each utterance lasted for about 4.1 and 2.3 seconds on average.
The \emph{RT03} data set consisted of two subsets: \emph{Fisher (FSH)} and \emph{SWBD}. Both of them contained 72 speakers. Approximate 2.5 and 2.7 minutes of speech data were recorded from each speaker respectively on average. The data amount for each speaker in the evaluation sets was thus significantly less than that in the training set.

\subsection{Experiments on 300-Hour Switchboard Setup}
\label{sec:300hr}

\subsubsection{\bf TDNN Baseline System}
\label{sec:baseline}

The baseline TDNN system was built mainly following the recipe\footnote{\em egs/swbd/s5c/local/chain/tuning/run\_tdnn\_7q.sh} of the Kaldi tookit~\cite{povey2011kaldi}. The network consisted of 15 context-splicing layers with 1536 nodes per layer. A 160-dimensional factored linear projection was employed prior to affine transformation in each context-splicing layer other than the first one. The ReLU function was used in the context-splicing layer, followed by batch normalization and dropout operations. The hidden layer input prior to context splicing was scaled and added to the hidden layer output by a skip connection. The splicing indices $\{-1,0,1\}$ and $\{-3,0,3\}$ were employed in the first 4 and the last 10 context-splicing layers respectively. After two 256 dimensional linear projections and batch normalization operations, the output layer generated the probabilities of 4456 HMM states.

For TDNN parameter optimization, the LF-MMI criterion~\cite{Povey2016Purely} was applied together with the minimum CE regularization. The back-propagation update was based on natural gradient~\cite{povey2014parallel}. Different from the recipe, 40-dimensional static filter-bank features were transformed by linear discriminative analysis to produce the input features. Speed perturbation was not used in this section. Due to the limited computing resource, the baseline TDNN was trained with 4 epochs of iterations on single job with one TITAN X (Pascal) GPU. No parallel models combination was used in each iteration.

In the decoding process, each utterance was first decoded by a 3-gram language model (LM) with a 30K word vocabulary which was built with the transcripts of training data. Lattice-based rescoring was then applied using a 4-gram LM built with the transcripts of both the \emph{Switchboard} training data and the \emph{Fisher English corpora}  (LDC2004T19 \& LDC2005T19) to obtain the best recognition result.

It has been shown that i-vector and LHUC adaptation could be jointly used to achieve complementary performance improvement in state-of-the-art ASR systems~\cite{luscher2019rwth}. Hence i-vector based adaptive training was used by the baseline TDNN system to investigate the complementary effect between i-vector and Bayesian adaptation.
The i-vector extractor was built by exploiting the first 100,000 utterances of training data.

Significance test based on the matched pairs sentence-segment word error was applied with the NIST SCTK~\cite{fiscus2007speech}. Two different models given the same evaluation set were utilized for the test, by setting the null hypothesis as no difference of average segmental errors was yielded between the two models~\cite{gillick1989some}. The significance level is defined as $P \leq 0.05$.

\subsubsection{\bf Analysis of Bayesian Adaptation Configurations}
\label{sec:bayes}

This section investigates the appropriate configurations of Bayesian adaptation approaches including the BLHUC, BHUB, and BPAct adaptation.
Table \ref{tab:config} shows the performance of applying Bayesian adaptation with different configurations to TDNN systems without using i-vector on the \emph{Hub5'00} data set. All speaker level data was used as adaptation data in this investigation.
For each update, the mini-batch size was 64.
The learning rates for LHUC and PAct were set to 0.01, and that for HUB was 0.000001, which were not normalized by the mini-batch size. All Bayesian adapted systems used the same random seed for fair comparison. The systems applying Bayesian adaptation consistently outperformed the corresponding systems applying deterministic adaptation. 

\begin{table}[!htbp]
  \begin{center}
  {
    \caption{Performance of applying deterministic or Bayesian adaptation with different configurations to the 300-hour TDNN system on the \emph{Hub5'00} data set. No i-vector was used in these systems. The results in bolt fonts are the lowest WERs in the corresponding columns. }
    \label{tab:config}
    \scalebox{0.72}[0.8]
    {
    \begin{tabular}{c|l|l|l|l|l|l|c|c|c} \toprule[0.7pt]\midrule[0.7pt]
           & \multirow{1}{*}{Adapt} & Adapt & Activation & Tied & \multirow{2}{*}{$J_{\text{est}}$} & Method of & \multicolumn{3}{c}{WER (\%) of \emph{Hub5'00}} \\ \cline{8-10}
           & \multirow{1}{*}{method} & \#layers & $\xi(\cdot)$ & var & & inference & \emph{SWBD} & \emph{CHE} & Avg. \\ \midrule[0.7pt]\midrule[0.7pt]
          (1) & -                   & -        & - & - & - & - & 10.1 & 20.6 & 15.4 \\ \midrule[0.7pt]
          (2) &\multirow{3}{*}{LHUC} & \multirow{3}{*}{1} & Identity & - & - & - & 9.7 & 19.3 & 14.6 \\
          (3) &                      & & 2Sigmoid & - & - & - & 9.8 & 19.2 & 14.6 \\
          (4) &                      & & Exponential    & - & - & - & 9.7 & 19.4 & 14.7 \\ \hline
          (5) &\multirow{6}{*}{BLHUC} & \multirow{6}{*}{1} & \multirow{9}{*}{Identity} & $\surd$ & 1 & Expectation & 9.6 & 18.7 & 14.2 \\
          (6) &                      & &  & $\times$  & 1 & Expectation  & 9.6 & 18.8 & 14.2 \\
          (7) &                      & &  & $\surd$ & 3 & Expectation  & 9.6 & 18.7 & 14.2 \\
          (8.1) &                      & &  & $\surd$ & 3 & $J_{\text{inf}}=3$  & 9.7 & 18.8 & 14.3 \\
          (8.2) &                      & &  & $\surd$ & 3 & $J_{\text{inf}}=8$  & 9.4 & 18.9 & 14.2 \\
          (8.3) &                      & &  & $\surd$ & 3 & $J_{\text{inf}}=16$  & 9.5 & 18.7 & 14.2 \\
          (8.4) &                      & &  & $\surd$ & 3 & $J_{\text{inf}}=32$  & 9.5 & 18.9 & 14.3 \\
          (8.5) &                      & &  & $\surd$ & 3 & $J_{\text{inf}}=64$  & 9.5 & 18.8 & 14.2 \\
          (8.6) &                      & &  & $\surd$ & 3 & $J_{\text{inf}}=128$  & 9.5 & 18.8 & 14.2 \\ \cline{4-10}
          (9) &                      & & 2Sigmoid & $\surd$ & 1 & Expectation & 9.7 & 19.0 & 14.4 \\
          (10) &                      & & Exponential    & $\surd$ & 1 & Expectation & 9.7 & 19.1 & 14.5 \\ \hline
          (11) & \multirow{3}{*}{LHUC} & \multirow{3}{*}{14} & Identity & - & - & - & 9.9 & 19.8 & 14.9 \\
          (12) &                      & & 2Sigmoid & - & - & - & 9.7 & 19.1 & 14.5 \\
          (13) &                      & & Exponential    & - & - & - & 9.9 & 20.0 & 15.0 \\ \hline
          (14) & \multirow{3}{*}{BLHUC} & \multirow{3}{*}{14} & Identity & $\surd$ & 1 & Expectation & {\bf 9.3} & {\bf 17.6} & {\bf 13.5} \\
          (15) &                      & & 2Sigmoid & $\surd$ & 1 & Expectation & 9.4 & 18.0 & 13.8 \\
          (16) &                      & & Exponential    & $\surd$ & 1 & Expectation & 9.8 & 18.9 & 14.4 \\ \midrule[0.7pt]\midrule[0.7pt]
          (17) & \multirow{2}{*}{HUB} & \multirow{2}{*}{1} & Identity & - & - & - & 9.7 & 18.9 & 14.4 \\
          (18) &                       &                       & Tanh & - & - & - & 9.6 & 18.9 & 14.4 \\ \hline
          (19) & \multirow{2}{*}{BHUB} & \multirow{2}{*}{1} & Identity & $\surd$ & 1 & Expectation & 9.5 & 18.8 & 14.2 \\
          (20) &                       &                       & Tanh & $\surd$ & 1 & Expectation & 9.5 & 18.8 & 14.3 \\ \hline
          (21) & \multirow{1}{*}{HUB} & \multirow{1}{*}{14} & Tanh & - & - & - & 9.5 & 18.7 & 14.3 \\ \hline
          (22) & \multirow{1}{*}{BHUB} & \multirow{1}{*}{14} & Tanh & $\surd$ & 1 & Expectation & {\bf 9.3} & 17.9 & 13.7 \\ \midrule[0.7pt]\midrule[0.7pt]
          (23) & \multirow{1}{*}{PAct} & \multirow{1}{*}{1} & - & - & - & - & 9.8 & 19.6 & 14.8 \\
          (24) & \multirow{1}{*}{BPAct} & \multirow{1}{*}{1} & - & $\surd$ & 1 & Expectation & 9.4 & 18.5 & 14.0 \\ \hline
          (25) & \multirow{1}{*}{PAct} & \multirow{1}{*}{14} & - & - & - & - & 10.2 & 21.1 & 15.7 \\
          (26) & \multirow{1}{*}{BPAct} & \multirow{1}{*}{14} & - & $\surd$ & 1 & Expectation & 9.5 & 18.3 & 14.0 \\ \midrule[0.7pt]\bottomrule[0.7pt]
    \end{tabular}
    }
  }
  \end{center}
\end{table}

The results shown by Systems (5) to (8.6) in Table \ref{tab:config} confirmed the combined use of tied variances, single sample in variational update ($J_{\text{est}} = 1$ in equation (\ref{eq:BLHUC:update0})) and taking the latent variable distribution expectation (equation (\ref{eq:bayes_infer_exp})) during inference, as suggested previously in Section \ref{sec:cost}, produced the best trade-off between computational efficiency and performance during Bayesian adaptation.
When the expectation in equation (\ref{eq:bayes_infer_exp}) was not used, a small number of samples in equation (\ref{eq:bayes_infer_mc}) was used such that the computational cost would not increase significantly.
Comparison between System (7) and Systems (8.1) to (8.6) show that, during inference, using the expectation is a good approximation and no statistically significant performance difference was observed when compared with drawing varying numbers of samples from 3 up to 128.
System (5) using $J_{\text{est}}=1$ (equation (\ref{eq:BLHUC:update0})), expectation for inference (equation (\ref{eq:bayes_infer_exp})), and tied variances made the adaptation simpler and more efficient while producing performance comparable to Systems (6) to (8.6). Hence this configuration 
would be used by other Bayesian adapted systems in the rest of this paper.

\begin{figure}[!htbp]
  \begin{center}
    \includegraphics[width=8.9cm]{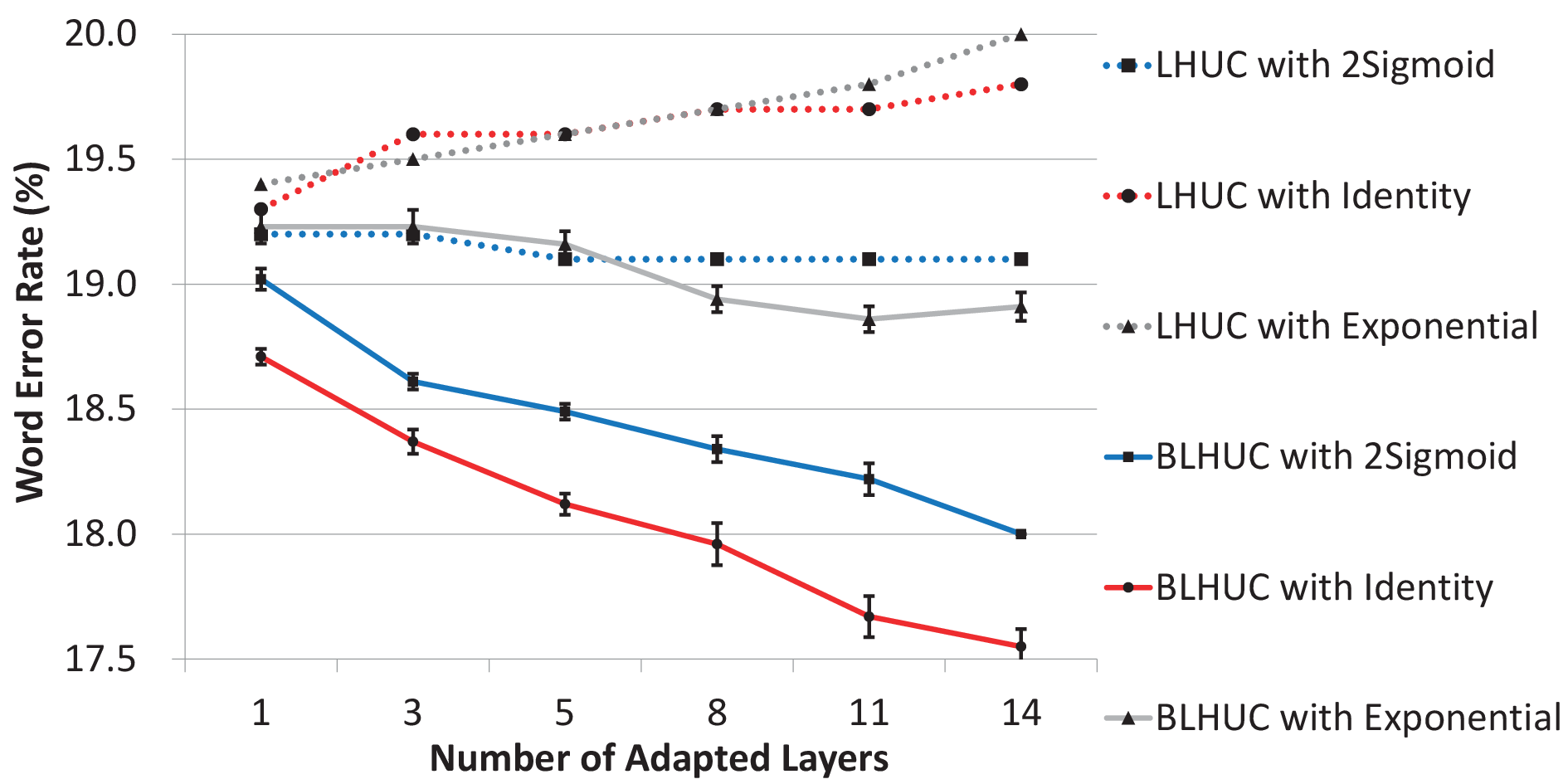}
    \caption{Performance contrast of applying LHUC and BLHUC adaptation to varying number of hidden layers and using different activation functions for LHUC scaling vectors, including 2Sigmoid, identity, and exponential. The evaluation was conducted on the \emph{CHE} subset of \emph{Hub5'00}. The BLHUC adaptation was repeated 10 times with different random seeds (for sampling in equation (\ref{eq:integal})) to show the average performance in word error rate (WER) and standard deviation. }
    \label{fig:activation}
  \end{center}
\end{figure}

Two trends could be observed from comparison between LHUC and BLHUC adapted systems (Systems (2) to (16) in Table \ref{tab:config}). The first trend suggested that applying BLHUC adaptation to the first 14 layers (Systems (14) to (16)) achieved better performance than to the first layer only (Systems (5) to (10)), while no significant difference was obtained by applying LHUC adaptation to varying numbers of layers. Second, using the identity activation as $\xi$ (in the left part of figure \ref{fig:adapt_bl}) achieved the best performance of BLHUC adapted to different numbers of layers (Systems (5) and (14)), while 2Sigmoid was the best one for standard LHUC.
Figure~\ref{fig:activation} shows more details of the trends on the \emph{CHE} data of \emph{Hub5'00} with systems applying LHUC or BLHUC adaptation to varying number of layers and using different activation functions for LHUC scaling vectors. The adapted layers were ranging from the first 1 to the first 14 layers.
In this figure the BLHUC adaptation was repeated 10 times with different random seeds (for sampling in equation (\ref{eq:integal})) to show the average performance in WER and standard deviation.
The performance of BLHUC adaptation went better when more layers were adapted, while more adapted layers showed overfitting in LHUC adaptation. This demonstrated the robustness of BLHUC adaptation against parameter uncertainty that linearly increased with respect to the number of deterministic SD parameters in the LHUC adapted systems.

Similarly, applying BHUB adaptation to 14 layers (Systems (22) in Table \ref{tab:config}) achieved better performance than to the first layer only (Systems (19) and (20)), while no significant difference was obtained by applying HUB adaptation to varying number of layers.
Moreover, using identity or Tanh activation functions as $\xi$ (in the middle part of figure \ref{fig:adapt_bl}) obtained similar performance, hence Tanh activations would be used for HUB/BHUB adaptation.
In contrast, no significant improvement was obtained by applying BPAct adaptation to more layer than only the first layer (Systems (24) v.s. (26)), while significant performance degradation resulted from applying PAct adaptation to 14 layers. This also demonstrated the robustness of BPAct adaptation against parameter uncertainty that linearly increased with respect to the number of deterministic SD parameters in the PAct adapted systems.


\subsubsection{\bf Robustness of Adaptation Using Varying Amounts of Speaker Data}
\label{sec:amount}

The robustness of Bayesian adaptation against varying amounts of adaptation data is investigated in this section. Table~\ref{tab:amount} shows the TDNN systems adapted with various amounts of adaptation data ranging from the first 5, 10, 20, 40, to all utterances of the \emph{CHE} subset of \emph{Hub5'00}. The aforementioned best configurations in Table \ref{tab:config} were used for the adaptation approaches (i.e., tied variances, $J_{\text{est}} = 1$ (equation (\ref{eq:BLHUC:update0})), and expectation (equation (\ref{eq:bayes_infer_exp})) for inference were used for all Bayesian adapted systems; LHUC/BLHUC adaptation with identity activations were applied to the first 14 layers; HUB/BHUB adaptation with Tanh activations and were applied to the first 14 layers; PAct/BPAct adaptation were applied to the first layer).
it could be seen that the Bayesian adapted systems consistently outperformed the systems adapted with deterministic SD parameters.
Moreover, using only 5 utterances (11 seconds on average) as adaptation data for Bayesian test-time adaptation led to significant absolute WER reduction of 0.7\% against the corresponding deterministic parameter adapted systems (Systems (3) v.s. (2), Systems (5) v.s. (4), and Systems (7) v.s. (6) in Table~\ref{tab:amount} respectively).

\begin{table}[!htbp]
  \begin{center}
  {
    \caption{Performance of applying Bayesian or deterministic adaptation approaches to 300-hour TDNN systems using varying amounts of adaptation data ranging from the first 5, 10, 20, 40, to all utterances of the \emph{CHE} subset in \emph{Hub5'00}. ``\dag'' means that the Bayesian adaptation significantly outperformed ($P \leq 0.05$) the corresponding deterministic adaptation. The results in bolt fonts are the lowest WERs in the corresponding columns. }
    \label{tab:amount}
    \scalebox{0.95}[0.95]
    {
    \begin{tabular}{c|l|c|c|c|c|c|c} \toprule[0.7pt]\midrule[0.7pt]
           & Adapt & \multirow{2}{*}{i-vec} & \multicolumn{5}{c}{WER (\%) w.r.t. \#Adapt utterances} \\ \cline{4-8}
           & method & & 5 & 10 & 20 & 40 & all \\ \midrule[0.7pt]\midrule[0.7pt]
          (1) & -   & $\times$ & 20.6 & 20.6 & 20.6 & 20.6 & 20.6 \\ \midrule[0.7pt]
          (2) & LHUC & \multirow{6}{*}{$\times$} & 19.7 & 19.2 & 19.4 & 19.3 & 19.8 \\
          (3) & BLHUC &  &  19.0\dag &  18.9 &  18.7\dag &  17.8\dag & 17.6\dag \\ \cline{1-2} \cline{4-8}
          (4) & HUB &  & 19.9 & 19.5 & 19.1 & 18.7 & 18.7 \\
          (5) & BHUB &  &  19.2\dag &  18.9\dag &  18.8 &  18.0\dag & 17.9\dag \\ \cline{1-2} \cline{4-8}
          (6) & PAct &  & 19.9 & 19.8 & 19.6 & 19.4 & 19.6 \\
          (7) & BPAct &  &  19.2\dag &  19.4\dag &  19.2 &  18.8\dag & 18.5\dag \\ \midrule[0.7pt]\midrule[0.7pt]
          (8) & -   & $\surd$ & 19.2 & 19.2 & 19.2 & 19.2 & 19.2 \\ \midrule[0.7pt]
          (9) & LHUC & \multirow{6}{*}{$\surd$} & 19.1 & 18.7 & 18.5 & 18.3 & 18.8 \\
          (10) & BLHUC &  &  18.1\dag &  17.7\dag &  17.3\dag &  {\bf 16.6}\dag & 16.7\dag \\ \cline{1-2} \cline{4-8}
          (11) & HUB &  & 19.1 & 18.6 & 18.5 & 18.0 & 17.8 \\
          (12) & BHUB &  &  {\bf 17.8}\dag &  {\bf 17.5}\dag &  {\bf 17.2}\dag &  17.1\dag & 16.9\dag \\ \cline{1-2} \cline{4-8}
          (13) & PAct &  & 19.5 & 19.3 & 18.6 & 18.3 & 18.4 \\
          (14) & BPAct &  &  18.1\dag &  18.4\dag &  18.3 &  17.6\dag & 17.5\dag \\ \midrule[0.7pt]\midrule[0.7pt]
          (15) & SAT LHUC & \multirow{6}{*}{$\surd$} & 18.8 & 18.7 & 18.6 & 18.3 & 18.6 \\
          (16) & SAT BLHUC &  &  17.9\dag &  17.7\dag &  17.3\dag &  16.9\dag &  {\bf 16.6}\dag \\ \cline{1-2} \cline{4-8}
          (17) & SAT HUB &  & 20.4 & 19.7 & 19.4 & 18.0 & 17.8 \\
          (18) & SAT BHUB &  & 19.9\dag & 19.2\dag &  18.6\dag &  17.5\dag & 17.1\dag \\ \cline{1-2} \cline{4-8}
          (19) & SAT PAct &  & 18.9 & 19.2 & 18.9 & 18.5 & 18.3 \\
          (20) & SAT BPAct &  &  17.9\dag &  18.2\dag &  17.6\dag &  17.7\dag & 17.5\dag \\ \midrule[0.7pt]\bottomrule[0.7pt]
    \end{tabular}
    }
  }
  \end{center}
\end{table}

When incorporating i-vectors and SAT, similar trends could be observed by using the Bayesian adaptation approaches.
When using only the first 5 utterances as adaptation data, the improvements achieved by Bayesian adaptation over deterministic adaptation were increased up to 1.4\% absolute (7.2\% relative) WER reduction (Systems (14) v.s. (13) in Table~\ref{tab:amount}). When using all speaker level data as adaptation data, the BLHUC adaptation (System (10)) significantly outperformed the LHUC adaptation (System (9)) by 2.1\% absolute (11.2\% relative) WER reduction.
However, when incorporating i-vectors, the performance of applying Bayesian adaptation to SAT systems (Systems (16), (18), and (20)) produced no significant improvement over the non-SAT systems (Systems (10), (12), and (14)). This indicated the similar effects of i-vector based adaptive training and applying model-based SAT.

\subsubsection{\bf Advantages over Other Regularized Adaptation Methods}
\label{sec:other}

This section compares the BLHUC adaptation to applying other regularizations to LHUC adaptation.
Table~\ref{tab:regular} shows the performance contrasts of these adapted TDNN systems without using i-vector. The regularization approaches included
f-MLLR (Systems (2) and (3) in Table~\ref{tab:regular}) which generated speaker-invariant input features and was shown to have good complementary effect to LHUC adaptation in \cite{swietojanski2016learning}, KL divergence regularization~\cite{yu2013kl} (System (4)),
lattice based supervision~\cite{padmanabhan2000lattice} (System (5))
which alleviated the sensitivity to errors in the supervision, MAP adaptation~\cite{huang2015maximum} (Systems (6) and (7)), and noisy LHUC adaptation (Systems (8) and (9)) which added noise sampled from the fixed distribution $p_0$ to the LHUC SD parameters. The noisy LHUC could be viewed as a special case of BLHUC with fixed variances and non-regularized update of SD distribution means.
Among these regularization approaches, all configurations were tuned to produce the best performance.
When using $\mathcal{N}(1,1)$ as the prior distribution, the MAP adaptation was simplified to L2 regularization~\cite{liao2013speaker}.
The empirical prior distribution was computed by equation (\ref{eq:prior_estimate}) in Section \ref{sec:prior}. 
The Table~\ref{tab:regular} shows that the BLHUC adapted systems (Systems (10) and (11)) consistently outperformed all the other regularization approaches. The noisy LHUC (Systems (8) and (9)) yielded comparable performance to BLHUC (Systems (10) and (11)) when the amount of adaptation data is small ($\leq 10$ utterances). This is expected as limited updates were performed in this setting and the estimated posterior in BLHUC would not diverge too far from the prior.
Moreover, no significant improvement was obtained by using an empirical prior (Systems (11)) against using $p_0 = \mathcal{N}(1,1)$ (Systems (10)) for BLHUC. 
Joint use of f-MLLR and BLHUC (System (12)) produced no improvement over using BLHUC alone.
\begin{table}[!htbp]
  \begin{center}
  {
    \caption{Performance contrasts of BLHUC adapted against other regularized LHUC adapted 300-hour TDNN systems using varying amounts of adaptation data evaluated on the \emph{CHE} subset of \emph{Hub5'00}. No i-vector was used in these systems. The results in bolt fonts are the lowest WERs in the corresponding columns. }
    \label{tab:regular}
    \scalebox{0.85}[0.9]
    {
    \begin{tabular}{c|l|l|c|c|c|c|c} \toprule[0.7pt]
          & \multirow{2}{*}{Adapt method} & \multirow{1}{*}{Prior $p_0$} & \multicolumn{5}{c}{WER (\%) w.r.t. \#Adapt utterances} \\ \cline{4-8}
          &                    &   & 5 & 10 & 20 & 40 & all \\ \midrule[0.7pt]
          (1) & LHUC                & - & 19.7 & 19.2 & 19.4 & 19.3 & 19.8 \\ \hline
          (2) & f-MLLR              & - & 27.3 & 24.2 & 21.3 & 19.8 & 19.5 \\
          (3) & f-MLLR + LHUC       & - & 27.1 & 23.8 & 21.0 & 19.6 & 19.0 \\ \hline
          (4) & KL regularized LHUC                & - & 19.6 & 19.3 & 19.2 & 19.2 & 19.8 \\ \hline
          (5) & Lattice based LHUC    & - & 19.9 & 19.9 & 19.7 & 19.3 & 19.2 \\ \hline
          (6) & \multirow{2}{*}{MAP LHUC} & $\mathcal{N}(1,1)$ & 19.6 & 19.3 & 19.2 & 19.1 & 19.4 \\
          (7) &                           & Empirical & 19.7 & 19.1 & 19.3 & 19.2 & 19.8 \\ \hline
          (8) & \multirow{2}{*}{Noisy LHUC ($\sim p_0$)} & $\mathcal{N}(0,0.5)$ & 19.2 & 18.9 & 19.0 & 18.8 & 19.6 \\
          (9) &                         & $\mathcal{N}(0,1)$   & 19.2 & 19.2 & 19.0 & 18.0 & 18.0 \\ \hline
          (10) & \multirow{2}{*}{BLHUC} & $\mathcal{N}(1,1)$ & {\bf 19.0} & 18.9 & 18.7 & 17.8 & 17.6 \\
          (11) &                           & Empirical & 19.4 & {\bf 18.8} & {\bf 18.4} & {\bf 17.7} & {\bf 17.5} \\ \hline
          (12) & f-MLLR + BLHUC & $\mathcal{N}(1,1)$ & 25.3 & 22.4 & 19.9 & 18.3 & 17.7 \\ \bottomrule[0.7pt]
    \end{tabular}
    }
  }
  \end{center}
\end{table}

\subsubsection{\bf Sensitivity of Adaptation to Different Learning Rates}

\begin{figure*}[!htbp]
  \begin{center}
    \includegraphics[width=18cm]{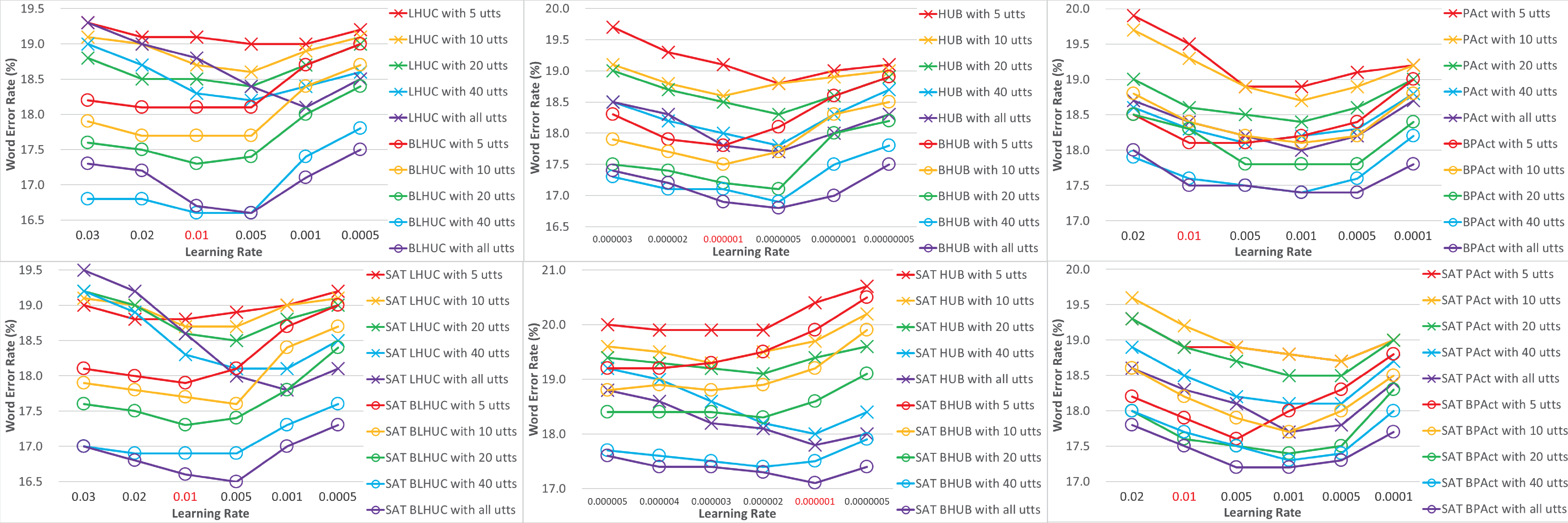}
    \caption{Performance in WER obtained by different adaptation techniques using different learning rates and various amounts of adaptation data, based on the systems using i-vector adaptation in Table \ref{tab:amount} (Systems (9) to (20)), including LHUC/BLHUC, HUB/BHUB, PAct/BPAct, and their SAT versions. The evaluation was conducted on the \emph{CHE} subset of \emph{Hub5'00}. The name of a curve, for example, ``LHUC with 5 utts'', means the results obtained by LHUC adaptation using 5 utterances as adaptation data. Among these curves, the ``$\times$'' denotes the results of deterministic adaptation techniques, the ``$\circ$'' denotes those of Bayesian adaptation techniques, and curves in different colour denote using different amounts of adaptation data. The empirically adjusted and fixed learning rates used for different adaptation methods (0.01 for LHUC/BLHUC and PAct/BPAct, and 0.000001 for HUB/BHUB) are denoted by the numbers in red colour on the ``Learning Rate'' axis. }
    \label{fig:learning_rate}
  \end{center}
\end{figure*}

In the previous sections, the learning rates of adaptation were empirically adjusted and fixed (i.e., set to 0.01 for LHUC/BLHUC and PAct/BPAct, and 0.000001 for HUB/BHUB) to prevent overtuning. Figure \ref{fig:learning_rate} shows the analysis about the effect of learning rate tuning on different adaptation techniques based on the systems using i-vector adaptation in Table \ref{tab:amount} (Systems (9) to (20)). It shows that the Bayesian adaptation consistently outperforms the comparable deterministic adaptation using different learning rates (by comparing curves of the same colour with ``$\times$'' and ``$\circ$'' legends in the sub-figure, representing each baseline deterministic adaptation method and its comparable Bayesian counterpart).

\subsection{Experiments on Data Augmented 900-Hour Switchboard Setup}
\label{sec:900hr}

In this section, speed perturbation on the rates of 0.9 and 1.1 was used to augment the 300-hour \emph{Switchboard} data such that three times of training data totalling 864 hours could be used. 
In addition to TDNN systems, a baseline CNN-TDNN system was built following the standard Kaldi recipe\footnote{\em egs/swbd/s5c/local/chain/run\_cnn\_tdnn\_1a.sh}. The CNN-TDNN employed 2-dimensional convolutional layers as the first 6 layers, where the two dimensions of the feature maps were the ``height'' and ``time'' axes. 
The input$-$ouput sizes of these convolutional layers in form of ``filter$\times$height'' were 6$\times$40$-$64$\times$40$-$64$\times$40$-$128$\times$20$-$ 128$\times$20$-$256$\times$10$-$256$\times$10 respectively.
Intuitively at the first convolutional layer when using identity filter weights, the symbol ``filter'' represented the number of filters being used, while ``height'' denoted the spectral dimensional context.
Each filter had the size 3$\times$3 in form of ``height$\times$time''. 
ReLU activations and batch normalizations were utilized following the 2560-dimensional convolution outputs.
Other configurations of the CNN-TDNN system were the same as the TDNN systems.
The same configurations of the Bayesian and deterministic adaptation approaches as in Table~\ref{tab:amount} were used in this section, except that, for the adapted CNN-TDNN systems, both identity and 2Sigmoid were used as activation functions in LHUC/BLHUC adaptation, and the second to sixth convolutional layers would not be adapted. 

Moreover, two unidirectional (forward/backward) context based long short-term memory recurrent neural network language models (LSTM LMs)~\cite{sundermeyer2012lstm} were built following the standard Kaldi recipe\footnote{\em egs/swbd/s5c/local/rnnlm/run\_tdnn\_lstm\_1e.sh}.
The transcripts of both the \emph{Switchboard} training data and the \emph{Fisher English corpora} were encoded as 1024 dimensional embedding vectors to train the LSTM LMs. Each LSTM LM consisted of 2 unidirectional LSTM layers with 1024 cells. Projections were used inside the LSTMs to reduce the output dimensions to 512. A context splicing layer with ReLU activation was exploited before and after each LSTM layers. The splicing indices of the three context splicing layers were $\{-1,0\}$, $\{-3,0\}$, and $\{-3,0\}$. Finally, an affine transformation was used to generate the output embedding vectors.
For decoding, the two LSTM LMs were employed successively to rescore the lattices generated by the 4-gram LM.
The supervision labels used in both deterministic and Bayesian adaptation were generated by the 1-best paths of lattices rescored by the LSTM LMs.

\subsubsection{\bf Performance of Bayesian Adaptation on 900-Hour Perturbed Setup}
\label{sec:lstm}

Table~\ref{tab:sp} presents the results of adapted TDNN systems using varying amounts of adaptation data. Consistent improvements were achieved by the Bayesian adaptation approaches over the deterministic adaptation approaches. Among these results, the BLHUC adaptation (Systems (3) and (10) in Table~\ref{tab:sp}) obtained the best performance when using varying amounts of adaptation data.
With LSTM LM restoring, the BLHUC adapted TDNN system (System (10)) significantly outperformed the comparable LHUC adapted baseline system (System (9)) when using 10 utterances (22 seconds on average) of speech or more as the adaptation data.

\begin{table}[!htbp]
  \begin{center}
  {
    \caption{Performance of applying Bayesian adaptation to the data augmented 900-Hour TDNN system using varying amounts of adaptation data evaluated on the \emph{CHE} subset of \emph{Hub5'00}. The i-vectors were used in these systems. ``\dag'' means that the Bayesian adaptation significantly outperformed ($P \leq 0.05$) the corresponding deterministic adaptation. The results in bolt fonts are the lowest WERs in the corresponding columns. }
    \label{tab:sp}
    \scalebox{0.99}[0.99]
    {
    \begin{tabular}{c|l|c|c|c|c|c|c} \toprule[0.7pt]\midrule[0.7pt]
          & Adapt & LSTM & \multicolumn{5}{c}{WER (\%) w.r.t. \#Adapt utterances} \\ \cline{4-8}
          & method               & LM  & 5 & 10 & 20 & 40 & all \\ \midrule[0.7pt]\midrule[0.7pt]
          (1) & \multirow{1}{*}{-}      & \multirow{7}{*}{$\times$} & 17.7 & 17.7 & 17.7 & 17.7 & 17.7 \\ \cline{1-2} \cline{4-8}
          (2) & \multirow{1}{*}{LHUC}      &  & 17.5 & 17.3 & 17.3 & 17.2 & 17.3 \\
          (3) & \multirow{1}{*}{BLHUC}      & & 16.9\dag & 16.9 & 16.6\dag & 16.1\dag & 16.0\dag \\ \cline{1-2} \cline{4-8}
          (4) & \multirow{1}{*}{HUB}      & & 17.8 & 17.3 & 17.0 & 16.8 & 16.6 \\
          (5) & \multirow{1}{*}{BHUB}      &  & 17.6 & 17.0 & 16.9 & 16.4 & 16.0\dag \\ \cline{1-2} \cline{4-8}
          (6) & \multirow{1}{*}{PAct}      &  & 17.7 & 17.7 & 17.7 & 17.3 & 17.0 \\
          (7) & \multirow{1}{*}{BPAct}      &  & 17.7 & 17.6 & 17.2\dag & 16.7\dag & 16.4\dag \\ \midrule[0.7pt]\midrule[0.7pt]
          (8) & \multirow{1}{*}{-}           & \multirow{7}{*}{$\surd$}         & 15.4 & 15.4 & 15.4 & 15.4 & 15.4 \\ \cline{1-2} \cline{4-8}
          (9) & \multirow{1}{*}{LHUC}        &   & 15.0 & 14.9 & 14.8 & 14.8 & 14.9 \\
          (10) & \multirow{1}{*}{BLHUC}       &    & {\bf 14.8} & {\bf 14.4}\dag & {\bf 14.2}\dag & {\bf 14.0}\dag & {\bf 13.8}\dag \\ \cline{1-2} \cline{4-8}
          (11) & \multirow{1}{*}{HUB}      &    & 15.5 & 15.1 & 14.9 & 14.4 & 14.2 \\
          (12) & \multirow{1}{*}{BHUB}     &    & 15.2 & 14.7 & 14.6 & 14.1 & 13.9 \\ \cline{1-2} \cline{4-8}
          (13) & \multirow{1}{*}{PAct}       &    & 15.5 & 15.1 & 15.1 & 14.8 & 14.7 \\
          (14) & \multirow{1}{*}{BPAct}      &     & 15.4 & 15.3 & 15.0 & 14.1\dag & 13.9\dag \\ \midrule[0.7pt]\bottomrule[0.7pt]
    \end{tabular}
    }
  }
  \end{center}
\end{table}


Table~\ref{tab:tdnn_all} shows the results of applying Bayesian adaptation approaches to both TDNN and CNN-TDNN systems evaluated on the \emph{Hub5'00} and \emph{RT03} data. All speaker level data was used as adaptation data.
It can be found that the Bayesian learning approaches, i.e., BLHUC, BHUB, and BPAct adaptation consistently outperformed both the corresponding deterministic approaches, i.e., LHUC, HUB, and PAct adaptation, as well as the un-adapted baseline system. Especially on the \emph{RT03} set, all the improvements produced by the Bayesian adaptation over deterministic adaptation are significant.
In the CNN-TDNN systems, there was no significant difference between using 2Sigmoid activation and identity activation for BLHUC adaptation (Systems (17.1) v.s. (17.2)). Therefore, only the 2Sigmoid activation was used for LHUC/BLHUC adaptation in the CNN-TDNN systems in the rest of this paper.
When using LSTM LM rescoring, the BLHUC adapted systems significantly outperformed the comparable LHUC adapted systems by relative WER reduction up to 10\% (The Avg. WERs on \emph{RT03} in Systems (10) v.s. (9) in Table~\ref{tab:tdnn_all}).

\begin{table}[!htbp]
  \begin{center}
  {
    \caption{Overall performance of applying Bayesian adaptation to the data augmented 900-Hour TDNN and CNN-TDNN systems evaluated on the \emph{Hub5'00} and \emph{RT03} data sets. The i-vectors were used in these systems. The adapted systems denoted by ``LHUC''/``BLHUC'' used identity activation functions, and those denoted by ``LHUC*''/``BLHUC*'' used 2Sigmoid activation functions. ``\dag'' means that the Bayesian adaptation significantly outperformed ($P \leq 0.05$) the corresponding deterministic adaptation. The results in bolt fonts are the lowest WERs in the corresponding columns. }
    \label{tab:tdnn_all}
    \scalebox{0.75}[0.8]
    {
    \begin{tabular}{c|l|l|c|c|c|c|c|c|c} \toprule[0.7pt]\midrule[0.7pt]
          & \multirow{2}{*}{Model} & Adapt & LSTM & \multicolumn{3}{c|}{WER (\%) of \emph{Hub5'00}} & \multicolumn{3}{c}{WER (\%) of \emph{RT03}} \\ \cline{5-10}
          & & method               & LM & \emph{SWBD} & \emph{CHE} & Avg. & \emph{SWBD} & \emph{FSH} & Avg. \\ \midrule[0.7pt]\midrule[0.7pt]
          (1) & \multirow{14}{*}{TDNN} & \multirow{1}{*}{-}      & \multirow{7}{*}{$\times$} & 9.4 & 17.7 & 13.6 & 19.8 & 12.3 & 16.2 \\ \cline{1-1} \cline{3-3} \cline{5-10}
          (2) & & \multirow{1}{*}{LHUC}      &  & 9.3 & 17.3 & 13.4 & 19.4 & 12.0 & 15.9 \\
          (3) & & \multirow{1}{*}{BLHUC}      &  &  8.9\dag &  16.0\dag &  12.5\dag &  17.2\dag & 10.8\dag & 14.2\dag \\ \cline{1-1} \cline{3-3} \cline{5-10}
          (4) & & \multirow{1}{*}{HUB}      &  & 9.1 & 16.6 & 12.9 & 18.3 & 11.5 & 15.1 \\
          (5) & & \multirow{1}{*}{BHUB}      &  &  8.9 &  16.0\dag &  12.5\dag &  17.2\dag &  10.7\dag &  14.1\dag \\ \cline{1-1} \cline{3-3} \cline{5-10}
          (6) & & \multirow{1}{*}{PAct}      &  & 9.2 & 17.0 & 13.2 & 18.9 & 11.8 & 15.5 \\
          (7) & & \multirow{1}{*}{BPAct}      &  &  8.9\dag & 16.4\dag & 12.6\dag & 17.8\dag & 11.0\dag & 14.5\dag \\ \cline{1-1} \cline{3-10}
          (8) & & \multirow{1}{*}{-}      & \multirow{7}{*}{$\surd$}        & 7.9 & 15.4 & 11.7 & 17.2 & 10.4 & 14.0 \\ \cline{1-1} \cline{3-3} \cline{5-10}
          (9) & & \multirow{1}{*}{LHUC}      &  & 7.7 & 14.9 & 11.4 & 16.5 & 10.1 & 13.5 \\
          (10) & & \multirow{1}{*}{BLHUC}      &   &  7.5 &  13.8\dag &  10.7\dag & 15.0\dag &  9.1\dag & 12.2\dag \\ \cline{1-1} \cline{3-3} \cline{5-10}
          (11) & & \multirow{1}{*}{HUB}      &      & 7.5 & 14.2 & 10.9 & 15.7 & 9.6 & 12.8 \\
          (12) & & \multirow{1}{*}{BHUB}      &     &  7.6 & 13.9 & 10.8 &  14.6\dag &  9.1\dag &  11.9\dag \\ \cline{1-1} \cline{3-3} \cline{5-10}
          (13) & & \multirow{1}{*}{PAct}      &     & 7.7 & 14.7 & 11.2 & 16.2 & 9.9 & 13.2 \\
          (14) & & \multirow{1}{*}{BPAct}      &    &  7.5 & 13.9\dag & 10.8\dag & 15.2\dag &  9.1\dag & 12.2\dag \\ \midrule[0.7pt]\midrule[0.7pt]
          (15) & \multirow{14}{*}{CNN-} & \multirow{1}{*}{-}      & \multirow{8}{*}{$\times$} & 8.7 & 16.9 & 12.9 & 19.1 & 11.7 & 15.5 \\ \cline{1-1} \cline{3-3} \cline{5-10}
          (16.1) & \multirow{14}{*}{TDNN} & \multirow{1}{*}{LHUC}      &  & 8.6 & 16.5 & 12.6 & 18.4 & 11.2 & 15.0 \\
          (16.2) & & \multirow{1}{*}{LHUC*}      &  & 8.5 & 16.2 & 12.4 & 17.9 & 10.8 & 14.5 \\
          (17.1) & & \multirow{1}{*}{BLHUC}      &  &  8.4 &  15.7\dag &  12.1\dag &  17.0\dag &  10.4\dag &  13.8\dag \\
          (17.2) & & \multirow{1}{*}{BLHUC*}      &  &  8.3\dag &  15.7\dag &  12.0\dag &  17.2\dag &  10.3\dag &  13.9\dag \\ \cline{1-1} \cline{3-3} \cline{5-10}
          (18) & & \multirow{1}{*}{HUB}      &  & 8.4 & 16.0 & 12.2 & 17.9 & 10.7 & 14.5 \\
          (19) & & \multirow{1}{*}{BHUB}      &  & 8.4 & 15.8 & 12.2 & 17.5\dag & 10.6\dag & 14.2\dag \\ \cline{1-1} \cline{3-3} \cline{5-10}
          (20) & & \multirow{1}{*}{PAct}      & & 8.6 & 16.5 & 12.6 & 18.2 & 11.0 & 14.8 \\
          (21) & & \multirow{1}{*}{BPAct}      & & 8.5 & 16.0\dag & 12.3\dag &  17.2\dag & 10.5\dag & 14.0\dag \\  \cline{1-1} \cline{3-10}
          (22) & & \multirow{1}{*}{-}      & \multirow{7}{*}{$\surd$}    & 7.5 & 14.9 & 11.3 & 16.6 & 9.7 & 13.3 \\ \cline{1-1} \cline{3-3} \cline{5-10}
          (23) & & \multirow{1}{*}{LHUC*}      &    & 7.1 & 13.9 & 10.5 & 15.3 & 8.9 & 12.2 \\
          (24) & & \multirow{1}{*}{BLHUC*}      &   &  {\bf 6.9} &  {\bf 13.2}\dag &  {\bf 10.1}\dag &  {\bf 14.5}\dag &  {\bf 8.5}\dag &  {\bf 11.6}\dag \\ \cline{1-1} \cline{3-3} \cline{5-10}
          (25) & & \multirow{1}{*}{HUB}      &    & 7.1 & 13.7 & 10.4 & 15.1 & 8.9 & 12.1 \\
          (26) & & \multirow{1}{*}{BHUB}      &    & 7.1 & 13.6 & 10.4 & 14.9\dag & 8.7\dag & 11.9\dag \\ \cline{1-1} \cline{3-3} \cline{5-10}
          (27) & & \multirow{1}{*}{PAct}      &    & 7.1 & 14.2 & 10.7 & 15.5 & 9.1 & 12.4 \\
          (28) & & \multirow{1}{*}{BPAct}      &    & 7.2 & 13.7\dag & 10.5 & 14.7\dag & 8.7\dag & 11.8\dag \\ \midrule[0.7pt]\bottomrule[0.7pt]
    \end{tabular}
    }
  }
  \end{center}
\end{table}

\subsubsection{\bf Performance Contrasts with Latest ASR System Results}
\label{sec:adv}

In this section, a larger LSTM LM in each (forward/backward) direction was trained by doubling the cell number to 2048 and projection dimensionality to 1024. Dropout operation with 85\% retention was applied to the output nodes of each layer. In addition, the SpecAugment technique~\cite{park2019specaugment} configured as in~\cite{tuske2020single} was applied to CNN-TDNN training with 6 epochs.
Table~\ref{tab:cnn_compare} compares the results of our BLHUC adapted CNN-TDNN systems to the most recent results reported in the literature. The System (8) in Table~\ref{tab:cnn_compare} was the same baseline CNN-TDNN system as System (22) in Table~\ref{tab:tdnn_all} rescored by the larger LSTM LMs.
Systems (1) and (2) in Table~\ref{tab:cnn_compare} were the BLSTM hybrid systems without and with affine transformation for environment adaptation~\cite{kitza2019cumulative}. 
System (3) was the listen, attend and spell E2E system built with SpecAugment~\cite{park2019specaugment}.
Systems (4) to (6) were the LSTM based attention encoder-decoder E2E systems built with SpecAugment~\cite{tuske2020single}.
System (7) was the hybrid encoder-decoder system jointly using both subword and phoneme information and built with SpecAugment~\cite{wang2020investigation}.
By using BLHUC speaker adaptation on the \emph{Hub5'00} data, our CNN-TDNN systems (Systems (10) and (13)) outperformed the Systems (1) to (5), and obtained comparable performance to System (7).
On the \emph{RT03} set, our BLHUC adapted system trained with SpecAugment (System (13) in Table~\ref{tab:cnn_compare}) achieved the best performance with WERs of 13.4\%, 7.9\%, and 10.7\% on \emph{SWBD} subset, \emph{FSH} subset, and the whole of \emph{RT03} respectively. This best system (System (13)) outperformed the LSTM based attention encoder-decoder system (System (6)) by using only 1/20 the number of parameters of System (6), as well as the hybrid encoder-decoder system (System (7)).
\begin{table*}[!htbp]
  \begin{center}
  {
    \caption{Performance contrasts of out BLHUC adapted CNN-TDNN systems rescored by larger LSTM LMs against other state-of-the-art systems conducted on the 300-hour Switchboard setup. The overall WERs in ``()'' are not reported in the literature and are estimated using the subset WERs. The results in bolt fonts are the lowest WERs in the corresponding columns. }
    \label{tab:cnn_compare}
    \scalebox{0.9}[0.9]
    {
    \begin{tabular}{c|l|l|c|c|c|c|c|c} \toprule[0.7pt]
          & \multirow{2}{*}{System} & \multirow{2}{*}{\#Param} & \multicolumn{3}{c|}{WER (\%) of \emph{Hub5'00}} & \multicolumn{3}{c}{WER (\%) of \emph{RT03}} \\ \cline{4-9}
          & & & \emph{SWBD} & \emph{CHE} & Avg. & \emph{SWBD} & \emph{FSH} & Avg. \\ \midrule[0.7pt]
          (1) & SMBR BLSTM hybrid + i-vectors~\cite{kitza2019cumulative} & - & 6.7 & 14.7 & 10.7 & - & - & - \\
          (2) & + Affine transform based environment adaptation & - & 6.7 & 13.5 & 10.2 & - & - & - \\ \hline
          (3) & Listen, attend and spell network + SpecAugment~\cite{park2019specaugment} & - & 6.8 & 14.1 & (10.5) & - & - & - \\ \hline
          (4) & \multirow{3}{*}{LSTM based attention encoder-decoder + SpecAugment~\cite{tuske2020single}} & 28.5M & 7.4 & 14.6 & (11.0) & - & - & - \\
          (5) &  & 75.1M & 6.8 & 13.4 & (10.1) & - & - & - \\
          (6) &  & 280.1M & 6.4 & {\bf 12.5} & ({\bf 9.5}) & 14.8 & 8.4 & (11.7) \\ \hline
          (7) & Hybrid encoder-decoder + CTC + SpecAugment~\cite{wang2020investigation} & - & {\bf 6.3} & 13.3 & (9.8) & - & - & 11.4 \\ \midrule[0.7pt]
          (8) &  {\bf LF-MMI CNN-TDNN + i-vectors (ours)} & 15.1M & 7.1 & 14.1 & 10.6 & 15.9 & 9.4 & 12.8 \\
          (9) & + {\bf LHUC based speaker adaptation} & + 11.8k & 6.8 & 13.5 & 10.2 & 14.5 & 8.6 & 11.7 \\
          (10) & + {\bf BLHUC based speaker adaptation} & + 11.8k & 6.6 & 12.8 & 9.7 & 13.8 & 8.2 & 11.1 \\ \hline
          (11) & {\bf LF-MMI CNN-TDNN + i-vectors + SpecAugment (ours)} & 15.1M & 7.2 & 13.9 & 10.6 & 15.1 & 9.1 & 12.2 \\
          (12) & + {\bf LHUC based speaker adaptation} & + 11.8k & 6.8 & 13.0 & 9.9 & 14.0 & 8.3 & 11.3 \\
          (13) & + {\bf BLHUC based speaker adaptation} & + 11.8k & 6.7 & 12.7 & 9.7 & {\bf 13.4} & {\bf 7.9} & {\bf 10.7} \\ \bottomrule[0.7pt]
    \end{tabular}
    }
  }
  \end{center}
\end{table*}

\section{Discussion and Conclusions}
\label{sec:conclu}

This paper proposes a DNN speaker adaptation framework using full Bayesian learning to account for SD parameter uncertainty given limited speaker specific adaptation data. Three exemplar approaches of DNN based speaker adaptation including Bayesian learning of hidden unit contributions (BLHUC), Bayesian parameterized activation functions (BPAct), and Bayesian hidden unit bias vectors (BHUB) are obtained by applying the Bayesian learning framework to the approaches using deterministic parameters including LHUC, parameterized activation functions (PAct), and hidden unit bias vectors (HUB). Latent variable posterior distributions are learned for each speaker in all of these Bayesian adaptation methods. An efficient variational inference based approach benefited from the Monte Carlo parameter sampling and re-parameterization is used to estimate the posterior parameters. Experiments conducted on the 300-hour speed perturbed \emph{Switchboard} setup of LF-MMI factored TDNN/CNN-TDNN systems using i-vector speaker adaptive training suggest the proposed Bayesian adaptation consistently outperforms the deterministic adaptation on the \emph{NIST Hub5'00} and \emph{RT03} evaluation sets. The efficacy of the proposed Bayesian adaptation techniques is demonstrated in a comparison against the state-of-the-art performance obtained on the same task using most recent hybrid and end-to-end systems reported in the literature.
The best proposed BLHUC adapted CNN-TDNN system achieves the state-of-the-art performance with 10.7\% WER on the \emph{RT03} evaluation set conducted on this task.
Future works may focus on the non-variational inference of the Bayesian adaptation.


%

\appendices

\section{Codes of The Bayesian Adaptation Based on Kaldi Toolkit}

The codes of the Bayesian adaptation is available in https://github.com/XIEXurong/kaldi.git. They are written based on the Kaldi toolkit~\cite{povey2011kaldi}. Please cite the published version of this article in the TASLP (DOI: 10.1109/TASLP.2021.3084072) if you use these codes.

\section{Updated Results Using The Bayesian Adaptation}

This part shows some new results obtained after the article submitted.

\subsection{Bayesian Learning of Linear Hidden Network}

As mentioned in Section \ref{sec:review}, linear hidden networks (LHN)~\cite{gemello2007linear} is a widely used transform based adaptation approach. This part shows the performance of applying Bayesian learning to the LHN, which is call Bayesian LHN (BLHN). This experiment was conducted on the 300-hour Switchboard setup without using the i-vector based adaptive training, as shown in Table \ref{tab:LHN}. The LHN adaptation was applied following the 160-dimensional linear projection of the second layer of the TDNN system, and estimated a $160 \times 160$ dimensional transformation for each speaker. The prior distribution for both MAP LHN and BLHN was empirically estimated as in Section \ref{sec:other}. On the \emph{CHE} subset, applying Bayesian learning to LHN adaptation achieved significant WER reduction compared to the standard LHN and MAP LHN. This shows the effectiveness of the proposed Bayesian learning framework on transform based adaptation approach.

\begin{table}[!htbp]
  \begin{center}
  {
    \caption{Performance of linear hidden network (LHN) adaptation and its improvements using maximum a posteriori (MAP) adaptation~\cite{huang2015maximum} and Bayesian learning based adaptation. }
   \label{tab:LHN}
    \scalebox{0.99}[0.99]
    {
    \begin{tabular}{c|l|l|c|c|c} \toprule[0.7pt]
          & \multirow{2}{*}{Adapt method} & \multirow{2}{*}{Initialization}  & \multicolumn{3}{c}{WER (\%) of \emph{Hub5'00}} \\ \cline{4-6}
          &                    & & \emph{SWBD} & \emph{CHE} & Avg. \\ \midrule[0.7pt]
          (1) & LHN                & Random & 10.1 & 20.6 & 15.4 \\
          (2) & MAP LHN             & Random & 10.0 & 19.8 & 15.0 \\
          (3) & BLHN      & Random & 9.8 & 19.3 & 14.6 \\ \hline
          (4) & LHN                & Identity & 9.9 & 19.9 & 14.9 \\
          (5) & MAP LHN             & Identity & 9.9 & 19.7 & 14.8 \\
          (6) & BLHN      & Identity & {\bf 9.8} & {\bf 19.0} & {\bf 14.4} \\ \bottomrule[0.7pt]
    \end{tabular}
    }
  }
  \end{center}
\end{table}

\subsection{Results on Baseline Systems Using Parallel Models Combination}

In the published article, the baseline systems were trained on single job without parallel models combination in each iteration due to the limited computing resource. As a complement, this part shows the performance of the BLHUC adaptation applied to the TDNN and CNN-TDNN systems trained with parallel models combination in each iteration. The Table \ref{tab:tdnn_all_combine} is corresponding to the BLHUC results in Table \ref{tab:tdnn_all}. The default parallel setting in the Kaldi toolkit was used, where the baseline systems were trained with 6 epochs beginning with 3 jobs and ending with 16 jobs. It shows that the Bayesian adaptation can be used together with the parallel models combination to obtain further improvement. For this purpose, Table \ref{tab:cnn_compare_combine} shows the new results corresponding to  the Table \ref{tab:cnn_compare} with parallel models combination used in systems (8)--(13). It shows that the BLHUC adapted systems with models combination (systems (10) and (13)) achieve the best overall performance on both \emph{Hub5'00} and \emph{RT03} data sets.

\begin{table}[!htbp]
  \begin{center}
  {
    \caption{Overall performance of applying BLHUC to the data augmented 900-Hour TDNN and CNN-TDNN systems evaluated on the \emph{Hub5'00} and \emph{RT03} data sets. The i-vectors were used in these systems. }
    \label{tab:tdnn_all_combine}
    \scalebox{0.75}[0.8]
    {
    \begin{tabular}{l|l|c|c|c|c|c|c|c} \toprule[0.7pt]\midrule[0.7pt]
          \multirow{2}{*}{System} & Adapt & LSTM & \multicolumn{3}{c|}{WER (\%) of \emph{Hub5'00}} & \multicolumn{3}{c}{WER (\%) of \emph{RT03}} \\ \cline{4-9}
          & method               & LM & \emph{SWBD} & \emph{CHE} & Avg. & \emph{SWBD} & \emph{FSH} & Avg. \\ \midrule[0.7pt]\midrule[0.7pt]
          \multirow{6}{*}{TDNN} & \multirow{1}{*}{-}      & \multirow{3}{*}{$\times$} & 9.4 & 17.7 & 13.6 & 19.8 & 12.3 & 16.2 \\ \cline{2-2} \cline{4-9}
          & \multirow{1}{*}{LHUC}      &  & 9.3 & 17.3 & 13.4 & 19.4 & 12.0 & 15.9 \\
          & \multirow{1}{*}{BLHUC}      &  & 8.9 & 16.0 & 12.5 & 17.2 & 10.8 & 14.2 \\ \cline{2-9}
          & \multirow{1}{*}{-}      & \multirow{3}{*}{$\surd$}        & 7.9 & 15.4 & 11.7 & 17.2 & 10.4 & 14.0 \\ \cline{2-2} \cline{4-9}
          & \multirow{1}{*}{LHUC}      &  & 7.7 & 14.9 & 11.4 & 16.5 & 10.1 & 13.5 \\
          & \multirow{1}{*}{BLHUC}      &   & 7.5 & 13.8 & 10.7 & 15.0 & 9.1 & 12.2 \\ \hline
          \multirow{6}{*}{TDNN} & \multirow{1}{*}{-}      & \multirow{3}{*}{$\times$} & 8.6 & 16.8 & 12.8 & 18.1 & 11.2 & 14.8 \\ \cline{2-2} \cline{4-9}
          \multirow{6}{*}{+ {\bf Model Combine}} & \multirow{1}{*}{LHUC}      &  & 8.7 & 16.6 & 12.6 & 17.9 & 11.2 & 14.7 \\
          & \multirow{1}{*}{BLHUC}      &  & 8.2 &  15.5 & 11.9 &  15.6 & 10.2 &  13.0 \\ \cline{2-9}
          & \multirow{1}{*}{-}      & \multirow{3}{*}{$\surd$}        & 7.1 & 14.4 & 10.8 & 15.8 & 9.5 & 12.8 \\ \cline{2-2} \cline{4-9}
          & \multirow{1}{*}{LHUC}      &  & 7.2 & 14.1 & 10.7 & 15.3 & 9.3 & 12.4 \\
          & \multirow{1}{*}{BLHUC}      &   & 7.0 &  13.1 & 10.1 &  13.6 & 8.5 &  11.1 \\ \midrule[0.7pt]\midrule[0.7pt]
          \multirow{6}{*}{CNN-TDNN} & \multirow{1}{*}{-}      & \multirow{3}{*}{$\times$} & 8.7 & 16.9 & 12.9 & 19.1 & 11.7 & 15.5 \\ \cline{2-2} \cline{4-9}
          & \multirow{1}{*}{LHUC}      &  & 8.5 & 16.2 & 12.4 & 17.9 & 10.8 & 14.5 \\
          & \multirow{1}{*}{BLHUC}      &  & 8.3 & 15.7 & 12.0 & 17.2 & 10.3 & 13.9 \\ \cline{2-9}
          & \multirow{1}{*}{-}      & \multirow{3}{*}{$\surd$}    & 7.5 & 14.9 & 11.3 & 16.6 & 9.7 & 13.3 \\ \cline{2-2} \cline{4-9}
          & \multirow{1}{*}{LHUC}      &    & 7.1 & 13.9 & 10.5 & 15.3 & 8.9 & 12.2 \\
          & \multirow{1}{*}{BLHUC}      &   & 6.9 & 13.2 & 10.1 & 14.5 & 8.5 & 11.6 \\ \hline
          \multirow{6}{*}{CNN-TDNN} & \multirow{1}{*}{-}      & \multirow{3}{*}{$\times$} & 8.6 & 16.7 & 12.7 & 18.5 & 11.3 & 15.1 \\ \cline{2-2} \cline{4-9}
          \multirow{6}{*}{+ {\bf Model Combine}} & \multirow{1}{*}{LHUC}      &  & 8.4 & 16.0 & 12.3 & 17.4 & 10.6 & 14.2 \\
          & \multirow{1}{*}{BLHUC}      &  & 8.4 &  15.3 & 11.9 &  16.5 & 10.1 &  13.5 \\ \cline{2-9}
          & \multirow{1}{*}{-}      & \multirow{3}{*}{$\surd$}    & 7.5 & 14.3 & 10.9 & 16.0 & 9.5 & 12.9 \\ \cline{2-2} \cline{4-9}
          & \multirow{1}{*}{LHUC}      &    & 7.3 & 13.7 & 10.5 & 14.8 & 8.8 & 11.9 \\
          & \multirow{1}{*}{BLHUC}      &   & {\bf 7.2} & {\bf 13.3}  & {\bf 10.3} &  {\bf 14.0} & {\bf 8.4} &  {\bf 11.3} \\ \midrule[0.7pt]\bottomrule[0.7pt]
    \end{tabular}
    }
  }
  \end{center}
\end{table}

\begin{table*}[!htbp]
  \begin{center}
  {
    \caption{Performance contrasts of out BLHUC adapted CNN-TDNN systems rescored by larger LSTM LMs against other state-of-the-art systems conducted on the 300-hour Switchboard setup. The overall WERs in ``()'' are not reported in the literature and are estimated using the subset WERs.}
    \label{tab:cnn_compare_combine}
    \scalebox{0.9}[0.9]
    {
    \begin{tabular}{c|l|l|c|c|c|c|c|c} \toprule[0.7pt]
          & \multirow{2}{*}{System} & \multirow{2}{*}{\#Param} & \multicolumn{3}{c|}{WER (\%) of \emph{Hub5'00}} & \multicolumn{3}{c}{WER (\%) of \emph{RT03}} \\ \cline{4-9}
          & & & \emph{SWBD} & \emph{CHE} & Avg. & \emph{SWBD} & \emph{FSH} & Avg. \\ \midrule[0.7pt]
          (1) & SMBR BLSTM hybrid + i-vectors~\cite{kitza2019cumulative} & - & 6.7 & 14.7 & 10.7 & - & - & - \\
          (2) & + Affine transform based environment adaptation & - & 6.7 & 13.5 & 10.2 & - & - & - \\ \hline
          (3) & Listen, attend and spell network + SpecAugment~\cite{park2019specaugment} & - & 6.8 & 14.1 & (10.5) & - & - & - \\ \hline
          (4) & \multirow{3}{*}{LSTM based attention encoder-decoder + SpecAugment~\cite{tuske2020single}} & 28.5M & 7.4 & 14.6 & (11.0) & - & - & - \\
          (5) &  & 75.1M & 6.8 & 13.4 & (10.1) & - & - & - \\
          (6) &  & 280.1M & 6.4 & 2.5 & (9.5) & 14.8 & 8.4 & (11.7) \\ \hline
          (7) & Hybrid encoder-decoder + CTC + SpecAugment~\cite{wang2020investigation} & - & {\bf 6.3} & 13.3 & (9.8) & - & - & 11.4 \\ \midrule[0.7pt]
           (8) & {\bf LF-MMI CNN-TDNN + i-vectors + SpecAugment + Model Combine} ({\bf ours}) & 15.1M & 6.9 & 13.2 & 10.1 & 14.6 & 8.4 & 11.7 \\
           (9) & + {\bf LHUC based speaker adaptation} & + 11.8k & 6.6 & 12.5 & 9.6 & 13.6 & 8.1 & 11.0 \\
           (10) & + {\bf BLHUC based speaker adaptation} & + 11.8k & 6.5 & {\bf 12.2} & {\bf 9.4} & {\bf 12.8} & {\bf 7.6} & {\bf 10.3} \\
           \bottomrule[0.7pt]
    \end{tabular}
    }
  }
  \end{center}
\end{table*}

\subsection{Experiment on The 150-hour HKUST Task}
\label{appendix_hkust}

The published article conducted experiments only on the \emph{Switchboard} data. As a complement, this part shows the performance of LHUC/BLHUC adaptation on the 150-hour \emph{HKUST} Mandarin conversational telephone speech database (LDC2005S15, LDC2005T32). The TDNN and CNN-TDNN system were built with the same configurations as the \emph{Switchboard} setup. Parallel models combination, speed perturbation (to 450 hours), and i-vector adaptation were used for training. The systems were evaluated on data sets of development set (\emph{Dev}) released in 2014~\cite{liu2015cambridge}, NIST \emph{RT03S} (LDC2007S10), and 1997 NIST Hub5 Mandarin evaluation set~\cite{zhang2015parameterised} (\emph{EVAL97}). The LSTM LM was trained with the transcription of training data. Table \ref{tab:hkust} shows the performance of applying LHUC/BLHUC adaptation to the TDNN and CNN-TDNN systems.

\begin{table}[!htbp]
  \begin{center}
  {
    \caption{Performance of LHUC/BLHUC adaptation on the database of \emph{HKUST} Mandarin conversational telephone speech evaluated on the data sets of \emph{Dev}, \emph{RT03S}, and \emph{EVAL97}.}
    \label{tab:hkust}
    \scalebox{0.99}[0.99]
    {
    \begin{tabular}{c|l|l|c|c|c|c} \toprule[0.7pt]\midrule[0.7pt]
          & \multirow{2}{*}{Model} & Adapt & LSTM & \multicolumn{3}{c}{WER (\%)} \\ \cline{5-7}
          & & method               & LM & \emph{Dev} & \emph{RT03S} & \emph{EVAL97} \\ \midrule[0.7pt]\midrule[0.7pt]
          (1) & \multirow{6}{*}{TDNN} & \multirow{1}{*}{-}      & \multirow{3}{*}{$\times$} & 24.1 & 36.9 & 36.4 \\
          (2) & & \multirow{1}{*}{LHUC}      &  & 23.8 & 36.3 & 35.6 \\
          (3) & & \multirow{1}{*}{BLHUC}      &  & 23.0 & 35.2 & 34.2 \\ \cline{1-1} \cline{3-7}
          (4) & & \multirow{1}{*}{-}      & \multirow{3}{*}{$\surd$}        & 22.0 & 35.3 & 34.6  \\
          (5) & & \multirow{1}{*}{LHUC}      &  & 21.5 & 34.4 & 33.9 \\
          (6) & & \multirow{1}{*}{BLHUC}      &   & 21.0 & 33.7 & {\bf 32.6} \\ \midrule[0.7pt]\midrule[0.7pt]
          (7) & \multirow{6}{*}{CNN-} & \multirow{1}{*}{-}      & \multirow{3}{*}{$\times$} & 23.7 & 36.2 & 36.0 \\
          (8) & \multirow{6}{*}{TDNN} & \multirow{1}{*}{LHUC}      &  & 23.2 & 35.6 & 35.4 \\
          (9) & & \multirow{1}{*}{BLHUC}      &  & 22.8 & 35.0 & 34.7 \\  \cline{1-1} \cline{3-7}
          (10) & & \multirow{1}{*}{-}      & \multirow{3}{*}{$\surd$}    & 21.8 & 34.4 & 34.4 \\
          (11) & & \multirow{1}{*}{LHUC}      &    & 21.4 & 33.8 & 34.1 \\
          (12) & & \multirow{1}{*}{BLHUC}      &   & {\bf 20.9} & {\bf 33.3} & 33.2 \\ \midrule[0.7pt]\bottomrule[0.7pt]
    \end{tabular}
    }
  }
  \end{center}
\end{table}

\subsection{Experiment on The DementiaBank Pitt Elderly Speech}

This part shows the performance of LHUC/BLHUC adaptation on the \emph{DementiaBank Pitt} corpus~\cite{becker1994natural}. In the experiment 15.75 hours of training data with 9.72 hours elderly participant data and 6.03 hours investigator data was used after silence stripping, and 3.14 hours of test data with 1.93 hours elderly \emph{Participant} data and 1.21 hours \emph{Investigator} data was used. CNN-TDNN was trained with the same configuration of Appendix \ref{appendix_hkust}. Table \ref{tab:dbank} shows the performance of applying LHUC/BLHUC adaptation as well as SAT to the CNN-TDNN system.

\begin{table}[!htbp]
  \begin{center}
  {
    \caption{Performance of applying LHUC/BLHUC adaptation to CNN-TDNN system on the \emph{DementiaBank Pitt} corpus evaluated on the test sets of \emph{Participant} and \emph{Investigator}. }
    \label{tab:dbank}
    \scalebox{0.8}[0.8]
    {
    \begin{tabular}{c|c|l|c|c|c|c} \toprule[0.7pt]\midrule[0.7pt]
          & Speed & Adapt & \multirow{2}{*}{LHUC SAT} & \multicolumn{3}{c}{WER (\%)} \\ \cline{5-7}
          & perturbation & method               & & \emph{Participant} & \emph{Investigator} & Avg. \\ \midrule[0.7pt]\midrule[0.7pt]
          (1) & \multirow{6}{*}{$\times$} & \multirow{1}{*}{-}      & \multirow{3}{*}{$\times$} & 38.4 & 17.6 & 29.3 \\
          (2) & & \multirow{1}{*}{LHUC}      &  & 37.5 & 17.2 & 28.6 \\
          (3) & & \multirow{1}{*}{BLHUC}      &  & 36.9 & 17.0 & 28.2 \\ \cline{1-1} \cline{3-7}
          (4) & & \multirow{1}{*}{LHUC}      & \multirow{2}{*}{$\surd$} & 37.3 & 17.1 & 28.4 \\
          (5) & & \multirow{1}{*}{BLHUC}      & & 36.4 & 16.6 & 27.8 \\ \midrule[0.7pt]\midrule[0.7pt]
          (6) & \multirow{6}{*}{$\surd$} & \multirow{1}{*}{-}      & \multirow{3}{*}{$\times$} & 36.9 & 17.1 & 28.3 \\
          (7) & & \multirow{1}{*}{LHUC}      &  & 36.0 & 16.6 & 27.5 \\
          (8) & & \multirow{1}{*}{BLHUC}      &  & 35.5 & 16.4 & 27.2 \\  \cline{1-1} \cline{3-7}
          (9) & & \multirow{1}{*}{LHUC}      & \multirow{2}{*}{$\surd$} & 35.4 & 16.4 & 27.1 \\
          (10) & & \multirow{1}{*}{BLHUC}      &   & {\bf 34.9} & {\bf 16.3} & {\bf 26.8} \\ \midrule[0.7pt]\bottomrule[0.7pt]
    \end{tabular}
    }
  }
  \end{center}
\end{table}


\section*{Acknowledgment}

This research is supported by Hong Kong Research Grants Council GRF grant No. 14200218, 14200220, Theme-based Research Scheme T45-407/19N, Innovation \& Technology Fund grant No. ITS/254/19, PiH/350/20, Shun Hing Institute of Advanced Engineering grant No. MMT-p1-19,
National Natural Science Foundation of China (U1736202), and National Key R\&D Program of China (2020YFC2004100)

\ifCLASSOPTIONcaptionsoff
  \newpage
\fi



  \bibliographystyle{IEEEtran}
  \bibliography{adapt_lda}

%



%

\begin{IEEEbiography}[{\includegraphics[width=1in,height=1.25in,clip,keepaspectratio]{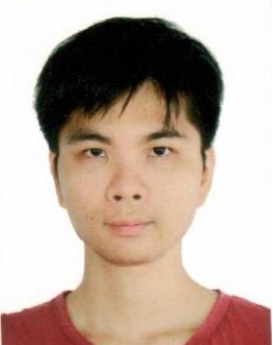}}]{Xurong Xie}
received the B.S. degree in mathematics and applied mathematics from Sun Yat-sen University, Guangzhou, China in 2011, the M.S. degree in computational statistics and machine learning from University College London, London, U.K. in 2012, and the Ph.D. degree in electronic engineering from The Chinese University of Hong Kong, Hong Kong, China in 2020. He is currently a Research Assistant Professor with the Shenzhen Institute of Advanced Technology, Chinese Academy of Sciences, Shenzhen, China. His research interests include machine learning, adaptation techniques for acoustic modeling, speech processing for disorder speech, and statistic methods applied to speech technologies such as Bayesian learning. He was the recipient of the Best Student Paper Award at IEEE ICASSP2019 for the paper titled ``BLHUC: Bayesian Learning of Hidden Unit Contributions for Deep Neural Network Speaker Adaptation''.
\end{IEEEbiography}

\begin{IEEEbiography}[{\includegraphics[width=1in,height=1.25in,clip,keepaspectratio]{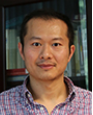}}]{Xunying Liu}
(Member, IEEE) received the undergraduate degree from Shanghai Jiao Tong University, Shanghai, China, and the M.Phil. degree in computer speech and language processing and the Ph.D. degree in speech recognition from the University of Cambridge, Cambridge, U.K. He was a Senior Research Associate with the Machine Intelligence Laboratory, Cambridge University Engineering Department, and since 2016, has been an Associate Professor with the Department of Systems Engineering and Engineering Management, The Chinese University of Hong Kong,
Hong Kong. His current research interests include machine learning, large vocabulary continuous speech recognition, statistical language modeling, noise robust speech recognition, audio-visual speech recognition, computer aided language learning, speech synthesis, and assistive technology. He is a Member of the ISCA. He and his students were the recipient of the number of best paper awards and nominations, including the Best Paper Award at ISCA Interspeech 2010 for the paper titled ``Language Model Cross Adaptation for LVCSR System Combination'', and the Best Paper Award at IEEE ICASSP2019 for their paper titled ``BLHUC: Bayesian Learning of Hidden Unit Contributions for Deep Neural Network Speaker Adaptation''.
\end{IEEEbiography}

\begin{IEEEbiography}[{\includegraphics[width=1in,height=1.25in,clip,keepaspectratio]{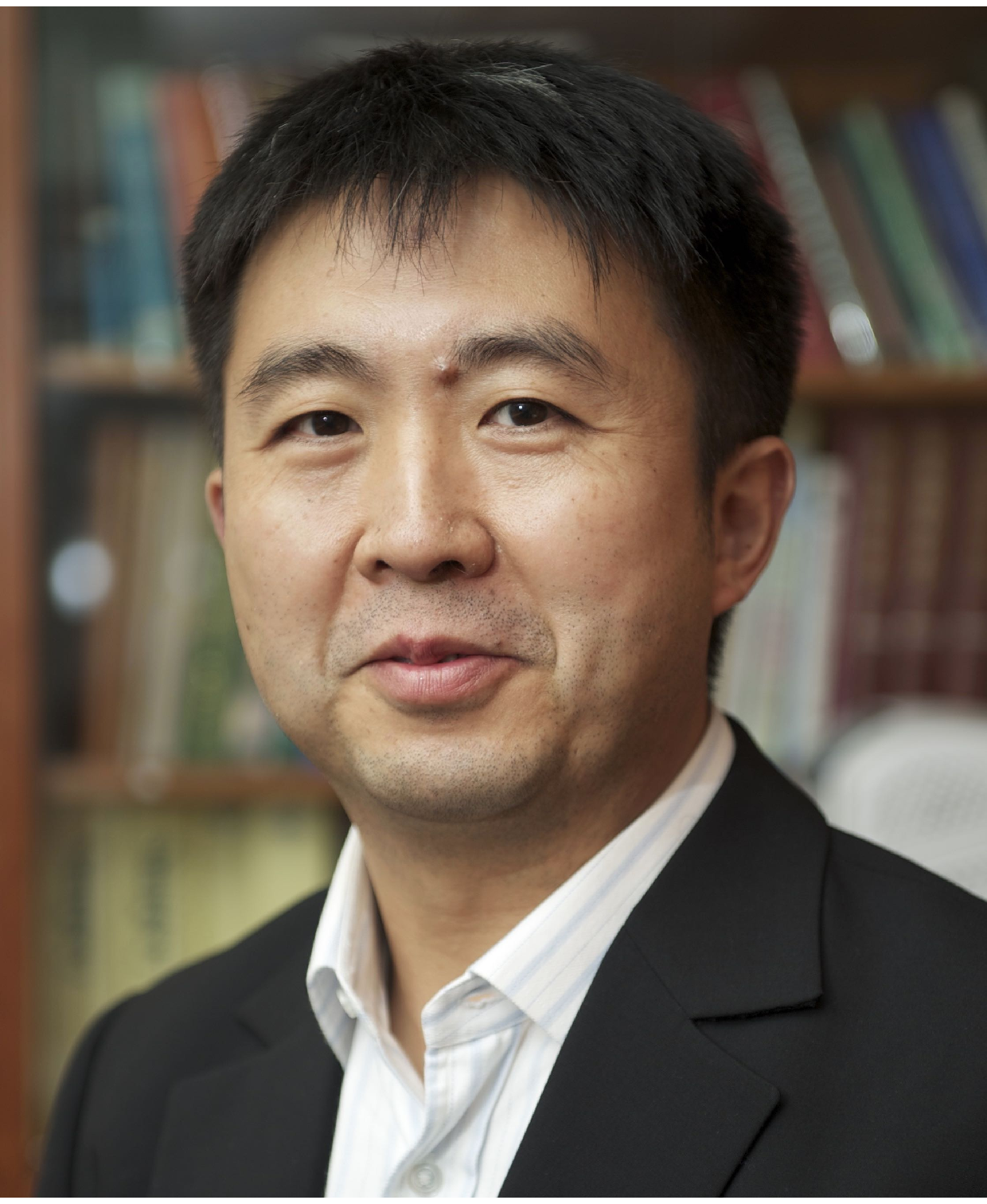}}]{Tan Lee}
(Member, IEEE) is an Associate Professor at the Department of Electronic Engineering and concurrently the Assistant Dean for Education at the Faculty of Engineering, the Chinese University of Hong Kong. His research covers speech and audio signal processing, automatic speech recognition, text-to-speech, paralinguistics in speech, and neurological basis of speech and language. Tan Lee led the effort on developing Cantonese-focused spoken language technologies that have been widely licensed for industrial applications. His recent work is focused on applying signal processing techniques and deep learning models to atypical speech in the context of low-resource languages and human communication disorders across different age groups. Tan Lee is an Associate Editor for the IEEE/ACM Transactions on Audio, Speech and Language Processing. He was an Area Chair of the Technical Programme Committees of INTERSPEECH 2014, 2016, and 2018, the General Co-Chair of ISCSLP 2018 and 2021. He is the Vice Chair of ISCA Special Interest Group on Chinese Spoken Language Processing.
\end{IEEEbiography}

\begin{IEEEbiography}[{\includegraphics[width=1in,height=1.25in,clip,keepaspectratio]{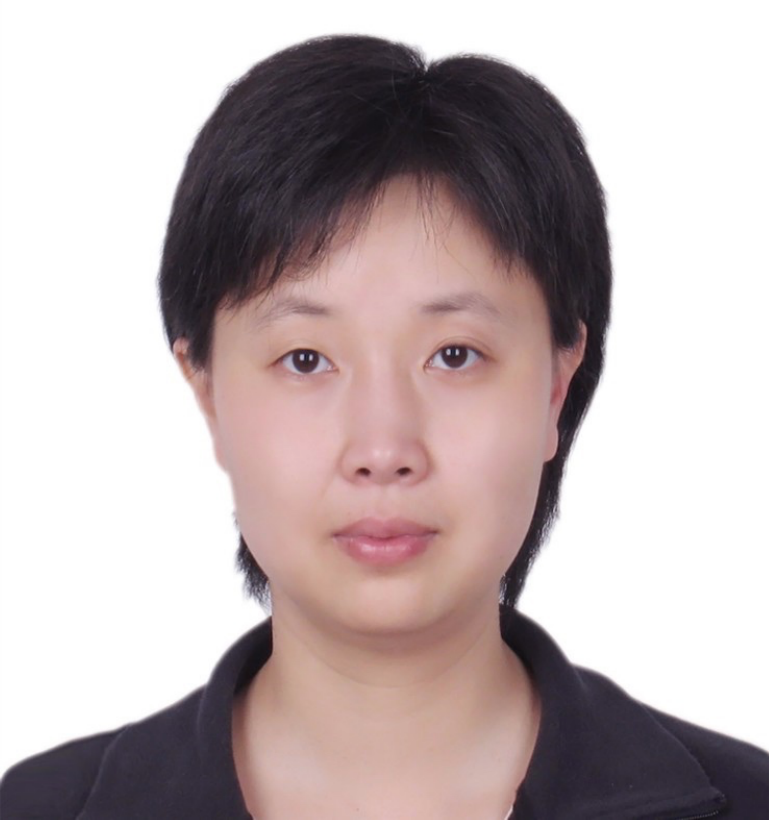}}]{Lan Wang}
(Member, IEEE) received the M.S. degree from the Center of Information Science, Peking University, Beijing, China, and the Ph.D. degree from the Machine Intelligence Laboratory, Department of Engineering, University of Cambridge, Cambridge, U.K. She worked on the autonomous Global Integrated Language Exploitation project funded under DARPAs Global Autonomous Language Exploitation program. She is currently a Research Professor with Shenzhen Institutes of Advanced Technology, Chinese Academy of Sciences, Shenzhen, China. Her research interests
include large vocabulary continuous speech recognition, speech visualization, and audio-visual speech processing.
\end{IEEEbiography}


%


\vfill
\vfill


\end{document}